\documentclass[acmsmall,screen,authorversion]{acmart}
\usepackage{multirow}
\AtBeginDocument{%
  }

\setcopyright{rightsretained}
\copyrightyear{2024}
\acmYear{2024}
\acmDOI{XXXXXXX.XXXXXXX}

\acmJournal{TOCHI}
\acmVolume{xx}
\acmNumber{x}
\acmArticle{xxx}
\acmMonth{5}

\acmISBN{978-1-4503-XXXX-X/18/06}

\begin{document}

\title[Self-Determination Theory and HCI Games Research]{Self-Determination Theory and HCI Games Research: \\
Unfulfilled Promises and Unquestioned Paradigms}

\author{April Tyack}

\author{Elisa D. Mekler}
\affiliation{%
  \institution{IT University of Copenhagen}
\country{Denmark}
}
\email{elme@itu.dk}

\renewcommand{\shortauthors}{Tyack and Mekler}

\begin{abstract}
Self-determination theory (SDT), a psychological theory of human motivation, is a prominent paradigm in human-computer interaction (HCI) research on games. However, our prior literature review observed a trend towards shallow applications of the theory. This follow-up work takes a broader view -- examining SDT scholarship on games, a wider corpus of SDT-based HCI games research (N=259), and perspectives from a games industry practitioner conference -- to help explain current applications of SDT. Our findings suggest that perfunctory applications of the theory in HCI games research originate in part from within SDT scholarship on games, which itself exhibits limited engagement with theoretical tenets. Against this backdrop, we unpack the popularity of SDT in HCI games research and identify conditions underlying the theory's current use as an oft-unquestioned paradigm. Finally, we outline avenues for more productive SDT-informed games research and consider ways towards more intentional practices of theory use in HCI.
\end{abstract}




\maketitle




%
\section*{In memory of April Tyack}
\label{memory}

 \emph{... our conceptual languages, though not necessarily at odds, were never identical to each other, were never seamlessly in harmony. 
 Indeed, though I’d say they were sympathetic languages by and large 
 ... %
 they stood somewhat at an angle to each other, 
 by turns converging and diverging in ways that, nevertheless, for better or worse, kept us talking to each other ... 
\vspace{0.2mm}
\begin{flushleft}How will I anticipate the give-and-take of clarifying response that not only enlivens real dialogue, real exchange, but enables real learning and real unlearning as well?
\end{flushleft} 
\vspace{0.5mm}
\begin{flushleft}So much is left suspended in the sudden, unlit absence ---
\end{flushleft}
\vspace{-2mm}
\begin{flushright}
\emph{Apology: On Intellectual Friendship \cite{scott2017stuart}}
\end{flushright}} 
\vspace{-3mm}

\section{Introduction}



Psychological concepts and models have long been employed in Human-Computer Interaction (HCI) to theorise the human user \citep{carroll1997human}. However, early applications of cognitive psychological theory did not develop into a coherent foundation of knowledge about human factors \citep{carroll2003introduction,clemmensen2006whatever,rogers2004new} -- circumstances that Rogers \citep[p.~22]{rogers2012hci} attributes to ``the stark differences between a controlled lab setting and the messy real world setting'' for which interactive artefacts and systems are designed. The deployment of broad theory in HCI has subsequently declined in the intervening years \citep{rogers2004new,rogers2012hci}, and this sporadic progress in theory development in domains such as usability and user experience (UX) has been identified as a cause for concern \citep{hornbaek2018commentary,kostakos2015big}. 


Research on games and play in HCI (henceforth HCI games research), however, 
has continued to employ broad psychological theories as foundational work \citep{poeller2022self,Tyack2016appeal}. One prominent example can be seen in self-determination theory (SDT) \citep{ryan2017self, ryan2019brick}, an influential theory of human motivation, which has provided HCI games research with propositions and concepts that can help explain motivational and experiential qualities of games and game-adjacent systems (e.g., gamification). 
In light of the theory's popularity, our previous work \citep{tyack2020self} investigated uses of SDT in games research at CHI and CHI PLAY, flagship venues for HCI games scholarship. 
In this prior review, we proposed that SDT is applied to HCI games research as a \emph{paradigm}; in other words, as ``a set of practices that a community has agreed upon, (including) questions to be asked and how they should be framed; phenomena to be observed, how findings from studies are to be analyzed and interpreted'' \citep[p.~4]{rogers2012hci}. We also observed a trend towards shallow applications of the theory, whereby SDT concepts and measures were used primarily as tools to evaluate the player experience (PX), 
with few deliberate attempts to apply the theory to explain, predict, or design for games and play phenomena. 

Although this prior work served as an effective overview of the domain, several lines of inquiry remain. First, our investigation of SDT in HCI games research referred only to work published at CHI and CHI PLAY. SDT-based work features prominently in these proceedings; however, the extent to which similar trends inhere within a wider games literature has remained uninvestigated. Second, 
we are yet to examine the ways that games and play are understood within SDT scholarship, i.e., as theorised by SDT co-developers and their colleagues. 
Studying these works may help explain current applications of the theory in HCI games research, and suggest further avenues of inquiry. Third, although our prior work provided an indication of the theory's relevance to HCI games research, questions remain regarding the extent to which SDT-based implications for design are applied by game developers themselves. This is particularly important to games scholars in HCI: industry outreach has consistently featured as a goal of the CHI games and play special interest group \citep{nacke2014games, nacke2016sigchi, nacke2018games, stahlke2020games}. Finally, although SDT's popularity in HCI games research is a given \citep{poeller2022self}, the reasons for this remain at best implicit. Interrogating SDT's popularity may thus point to conditions that facilitate theory uptake in HCI. 

This follow-up work consequently takes a broader view of SDT as applied in research and design practice, placing an expanded corpus (N=259) of work from more varied conference and journal venues in conversation with foundational SDT scholarship on games, and the work of games industry practitioners. Many of the issues identified in our prior review \citep{tyack2020self} are common to our expanded corpus: Perfunctory applications of SDT make up the bulk of the literature, and theoretical misconceptions are prevalent, even in works that offer more substantial engagement with the theory. Need satisfaction and intrinsic motivation remain the fundamental concepts used to investigate PX, gamification, and other central topics.  
Notably, the expanded corpus features more work with tighter theoretical links, where SDT is leveraged to inform the research direction, derive hypotheses, and interpret results. However, few attempts are made in these papers to extend or 
challenge the theory. Indeed, we identify a broad unwillingness to contest SDT tenets when study results are inconsistent with the theory -- indicating that SDT largely figures as an \emph{unquestioned paradigm} in HCI games research. 
Crucially, our analysis suggests that these issues originate in part from \emph{within} SDT scholarship on games, which itself exhibits a slew of theoretical inconsistencies, unfounded claims, underspecified propositions, and limited engagement with SDT mini-theories. 

Finally, our analysis of presentations (N=16) from the Game Developers Conference (GDC) -- a leading venue for game development knowledge-sharing -- reveals design practitioners' widespread familiarity with foundational SDT concepts, which suggests further opportunities for HCI games research to operate as a bridge between SDT and design practice. 
However, SDT-based HCI literature currently seems 
entirely %
absent %
in game practitioner discourse. 
Against this backdrop, 
we examine the conditions underlying SDT's popularity in games research and industry practice, and the ways these relate to the theory's current use as paradigm. 
%
To support more intentional practices of theory use, we 
identify opportunities for HCI games research to talk back to SDT and inform design practice, and locate productive avenues for SDT-based games research based on un(der)used areas of the theory. 
%
Beyond SDT, 
this work extends our understanding of the 
import and impact of theory in HCI \citep{girouard2019reality,rogers2004new,rogers2012hci}, especially 
with respect to 
why certain theories gain traction over others, and how the conditions that facilitate theory uptake 
shape the ways theory is put to use. 


A note on structure, acknowledging that readers will be interested in different aspects of the present work depending on their familiarity with SDT, as well as their investment in games and HCI scholarship: 
We first review the roles and uses of theory in HCI (\autoref{theory}), situating our prior 
work on 
SDT in HCI games research \citep{tyack2020self} amidst the wider HCI landscape. This is followed by an overview of SDT's core assumptions (\autoref{sdt}), key tenets and six mini-theories (sections \ref{cet}-\ref{rmt}). 
Next, we chronicle SDT games scholarship (\autoref{games_sdt}), that is, games research authored by SDT co-developer Richard Ryan and associates. This section is recommended also to readers with only passing interest in games, as it serves to contextualise many of the present work's findings and implications. We then report our expanded literature review of SDT-based HCI games research, with the methodological procedure and results described in \autoref{lit} and \autoref{Results}, respectively. Industry practitioner perspectives on SDT are presented in \autoref{gdc}. Finally, \autoref{Discussion} synthesises across our findings to help explain why HCI games research has predominantly applied SDT as an unquestioned paradigm and establish ways towards more intentional practices of theory use. We then consider avenues for more productive SDT-informed HCI games research  (\autoref{avenues}), followed by the present work's broader implications for theory in HCI (\autoref{hci}), as well as limitations and open questions for future work (\autoref{limitations}).

\section{Uses of Theory in Human-Computer Interaction}
\label{theory}
In this paper, we examine the different uses of SDT in games research and design practice -- but what is theory good for? HCI and adjacent fields have a long history of scholarship on the purposes and benefits of theory. 
The categorisations of HCI theory produced by 
Bederson and Shneiderman \citep{bederson2003theories} and Rogers \citep{rogers2012hci} 
have been particularly influential, and 
highlight that theory can (among other things) establish common terminology, explain and contextualise research findings, predict outcomes, inform design practice, and generate questions for further study. Halverson \citep{halverson2002activity} similarly identifies four attributes of theory in computer-supported collaborative work -- its powers of description, rhetoric, inference, and application (e.g., to design) -- and further elaborates ways that theories vary in their capacity for each. With regards to the role of theory for design, Oulasvirta and Hornbæk \citep{oulasvirta2022counterfactual} argue 
that theory serves as a `speculation pump', whereby a theory's constructive power is contingent on it offering counterfactual propositions that link design antecedents to desirable and reliable outcomes. Gaver \citep{gaver2012what}, in contrast, 
emphasises the generative potential of theory 
to inspire new designs. He further notes that theory underspecifies design practice ``by necessity'' \citep[p.~940]{gaver2012what}, but can instead help articulate and make apparent design considerations across a set of particular artefacts. Finally, Bardzell \citep{bardzell2009interaction} outlined different purposes that aesthetics and critical theory may serve in HCI, such as informing different stages of design, but also as a means to subvert and critique the design process and its consequences. 


A number of works \citep[e.g.,][]{dalsgaard2014between,hook2012strong,stolterman2010concept} have extolled forms of `intermediate-level knowledge' 
as a vehicle for translating 
theory between academic HCI and industry design practices. 
Velt et al.~\citep{velt2020translations}, for example, report on a 5-year project of translating the trajectories framework \citep{benford2009interaction} for use in UX design, a process that was driven as much by the designers as the research team. Importantly, the ways that designers interpreted and applied the trajectories framework diverged, and in some cases extended, its academic specification. For instance, the originally proposed `interface' component \citep{benford2009interaction} was split into `hardware' used to access content and the `commercial service' that provides it. In this way, flexible approaches to translation may 
facilitate the uptake of theory in design practice. 
Ploderer et al.~\citep{ploderer2021diagramming} instead present `field theories', another form of intermediate-level knowledge based in diagramming, which aim to form tighter links between research findings and artefact design. Crucially, this approach is iterative, involving one or more evaluation stages in which prototype designs are tested for congruence with the associated field theory, ensuring that the final results -- the field theory and artefact design alike -- appropriately represent, and respond to, the situation(s) of interest. Finally, Beaudouin-Lafon et al.~\citep{beaudouin-lafon2021generative} propose `generative theories of interaction', which operate between a core grounding theory (e.g., that concerns human activity and behaviour with technology) 
and artefact design practice. Like field theories, generative theories of interaction are iterated upon through the design process, and the results of this work may consequently influence or extend the grounding theory.

In parallel with scholarship that advocates for how HCI \emph{should} engage with theory, a smaller literature investigates the actual use of theories and frameworks in academic and industry practice. 
Clemmensen \citep{
clemmensen2005community}, for example, found that Danish usability professionals valued HCI theory (e.g., distributed cognition, situated action) for communicative purposes, to establish a common frame of reference or legitimise design decisions. In Rogers' study \citep{rogers2004new}, in contrast, few industry practitioners reported using theories from HCI in their work, 
not for lack of perceived utility, but because they did not know how to apply them effectively. This disconnect, in turn, has been attributed to 
HCI scholars' disregard for the realities of design practice \citep{colusso2017,gray2014reprio}. 

Speaking to 
theory use in academic practice, Hekler et al.~\citep{hekler2013} 
observed that HCI researchers employ behavioural theories \citep[e.g., the transtheoretical model,][]{prochaska2008} to guide the design and evaluation of behaviour change technologies. However, they also noted a tendency in the literature to cherrypick individual constructs \citep[e.g., contemplation,][]{prochaska2008} instead of leveraging theoretical tenets \citep[e.g., `higher self-efficacy increases the likelihood that an individual will transition from the contemplation to the behaviour preparation stage',][]{west2019development} -- which makes it difficult to reason about \emph{why} a technology is effective or not. Examining 
scholarship on artificial moral agents \citep{zoshak2021beyond}, Zoshak and Dew found that ethical theories are primarily leveraged to inform technical implementation and support analytical arguments. They further observed that the field seems to privilege deontological and consequentialist perspectives 
over alternate ethical frameworks. The authors note that these trends in theory use not only narrow the ways researchers and developers think about artificial agents, but
risk 
reinforcing existing inequalities by encoding 
hegemonic ethical norms into technology. 

Turning to specific theories, Clemmensen et al.~\citep{clemmensen2016making} used citation analysis to evaluate Activity Theory's influence, concluding that HCI had fruitfully applied the theory to inform design, as well as for guiding conceptual and empirical analyses of HCI phenomena, and in turn produced key reference texts used in other disciplines. Similarly, a review of the Reality-Based Interaction framework \citep{girouard2019reality} determined that the framework was most frequently cited to justify research decisions, inform the design of studies or novel artefacts, and support claims. Substantive applications of the framework appeared to be declining over time, however.  Taking papers rather than citations as the primary unit of analysis, Velt et al.~\citep{velt2017trajectories} identified five ways the trajectories framework \citep{benford2009interaction} had been applied in HCI -- to situate work, analyse and describe experience, influence design, and develop or critique concepts. Although papers in the latter category accounted for over half of their corpus, the authors observe that the vast majority of works extending the trajectories framework were co-authored by its original creators, suggesting a broader aversion to collective theory development. 

Finally, a recent analysis \citep{oulasvirta2022counterfactual} of CHI 2017 best papers observed few instances of deliberate theory use, with works rarely stating how design choices were informed by theory. 
Oulasvirta and Hornbæk \citep{oulasvirta2022counterfactual} claim that this limited theoretical engagement stems from iterative design practices and HCI researchers' scepticism towards the relevance of (scientific) theories for design. 
Beck and Stolterman \citep{beck2016examining} likewise critique perfunctory applications of theory in design research, arguing for more intentional practices of selecting and implementing theory, and greater attention to the ways that research findings can `talk back' to theoretical assumptions, i.e., how they relate to the research question, previous findings, or suggest changes to a particular theory. 

In the precursor to the present work \citep{tyack2020self}, 
we observed similar 
patterns of limited theory use with regards to the ways HCI games scholars applied SDT in their research: 
the majority of works 
cited SDT 
when situating their research or contextualising results, presenting implications for design, or selecting SDT-based self-report measures -- rarely with a theory-driven rationale. Few papers employed SDT as a basis for the research, for instance, to generate 
study hypotheses or inform design decisions, 
leaving ample potential to more fruitfully leverage its tenets and mini-theories (see \autoref{sdt} below). Although several works featured implicit assumptions about the impact of their designs on SDT-based concepts, few explicated why they expected these results. 
As such, we concluded that HCI games research presently figures SDT less as a theory, and more as a paradigm \citep{rogers2012hci} that structures games research by means of a shared vocabulary, a conceptualisation of (good) PX grounded in need satisfaction and intrinsic motivation, as well as the provision of measures -- the Player Experience of Need Satisfaction scale \citep[PENS;][]{rigby2007player} and the Intrinsic Motivation Inventory \citep[IMI;][]{centerIMI} -- to operationalise said concepts. 



\section{Self-Determination Theory}
\label{sdt}
Before we present our expanded analysis of the uses of SDT in games research and design practice, we provide here a brief primer covering the theory's core assumptions, as well as its key concepts and mini-theories. We revisit these in sections \ref{Discussion} and \ref{avenues} to discuss potential theoretical misconceptions and to identify as of yet untapped avenues for SDT-informed HCI games research. An exhaustive 
account of SDT is beyond scope, 
but we refer readers to \citep{ryan2017self,ryan2019brick,ryan2023oxford} for more detail.

%
SDT is a psychological macro-theory\footnote{To our knowledge, SDT authors have never explicitly specified their understanding of `macro-theory'. 
The term could refer to the theory's consideration of different levels of analysis ranging from the micro-level \citep[e.g., neurological correlates of intrinsic motivation][]{domenico2017emerging} to economic and political systems \citep{ryan2017economic,ryan2019brick}; its tenets being ``embedded in an organismic-dialectical metatheory'' \citep[p.~229]{deci2000what}; 
or that SDT spans multiple interrelated mini-theories \citep[see][]{barnard2000systems}. We surmise that usage of the term `macro-theory' in SDT scholarship pertains to all these meanings \citep[see also][p.~16]{koole2019becoming}.
} of human motivation, growth, and wellbeing \citep{deci2002handbook, ryan2017self, ryan2019brick,ryan2021building} that characterises humans as fundamentally active organisms. In particular, SDT posits \textit{intrinsic motivation}, an innate human propensity for inherently satisfying activity, and \textit{organismic integration}, which directs the assimilation and organisation of external stimuli into the developing self. Individuals are considered to `thrive' and experience wellbeing to the extent that their actions reflect the truest values of the self \citep{ryan2013flourishing}. Motivation, integration, and wellbeing are energised by the satisfaction of three basic psychological needs: \emph{competence}, the feeling of having an effect; \emph{autonomy}, a sense that actions are self-endorsed and performed willingly; and \emph{relatedness}, a sense of reciprocal care, value, and belonging in relation to other social figures and collectives \citep{ryan2017basic}.

At its core, SDT is considered a \emph{scientific} theory \citep{ryan2017self, ryan2019brick}, in that it contains a number of concepts (e.g., competence, intrinsic motivation) and empirically-testable propositions
that are assumed to generalise across varied contexts \citep{whetten1989constitutes}, and which serve to explain and predict the impact of certain events on motivation and wellbeing. Propositions are statements that specify the relation between theoretical concepts \citep{whetten1989constitutes}, for example, ``events that promote greater perceived competence enhance intrinsic motivation'' \cite[p.~130]{ryan2017cet}. In psychology, theories are distinguished from phenomena: empirically-backed research findings that have not been collated into a more general explanatory theory \citep{schlinger1998science}. SDT co-developers Ryan and Deci recapitulate these qualities as defining features of the theory. Specifically, in contrast with psychological models -- which ``bring specific phenomena into focus, but are rarely coordinated with each other, or with generalisable principles'' \citep[p.~113]{ryan2019brick}, 
they locate the utility of SDT in its status as a \emph{broad} theory 
that collates empirical findings into a general explanatory framework, which 
``
demands of itself clinical, qualitative, and conceptual critique, and must pass the criteria of epistemological coherence and rigour'' \citep[p.~113]{ryan2019brick}.

Finally, Ryan and Deci emphasise the theory's ``strong \emph{translational} value'' \citep[p.~112, emphasis added]{ryan2019brick} in that it is ``\emph{practical} insofar as it points to how features of contexts [...] facilitate or undermine motivation''
, as well as ``\emph{critical} insofar as it examines proximal social contexts [...] as well as more pervasive cultural, political, and economic conditions in terms of their adequacy in supporting versus impairing human thriving'' \citep[p.~4, emphases in original]{ryan2017introduction}. Consequently, SDT has been applied to a wide range of domains 
and interventions, such as education \citep{earl2017autonomy}, health \citep{ng2012health,teixeira2020classification}, parenting \citep{grolnick1989parent}, sports \citep{sheehan2018associations}, and the workplace \citep{gillet2015effects}. Likewise, SDT has risen to prominence in HCI and adjacent fields \citep{ballou2022self}, with applications in user experience \citep{bruhlmann2018UMI,hassenzahl2010needs,peters2018designing,zhao2023older}, information systems \citep{hornbaek2017technology,jia2021needs,rezvani2017motivating}, behaviour change technology \citep{gerstenberg2023designing,villalobos-zuniga2020apps}, human-AI interaction \citep{calvo2020supporting,xi2021designing}, as well as HCI games \citep{spiel2021purpose,tyack2020self} and gamification research \citep{krath2021revealing,seaborn2015gamification}. For these reasons, SDT has also been proposed as 
a potential ‘motor theme’ \citep{kostakos2015big} that could drive more cumulative and integrative research across seemingly disparate strands of HCI scholarship \citep{ballou2022self}.

SDT is currently organised into six mini-theories, whose underlying concepts are continuously developed, critiqued, and revised \citep[e.g.,][]{ryan2021legacy,vallerand2008reflections, vansteenkiste2010development}. In the following, we briefly describe the key tenets and mini-theories of SDT, in order of their historical development. We then outline key developments in SDT-based games research.

\subsection{Cognitive Evaluation Theory}
\label{cet}

\emph{Cognitive evaluation theory} (CET) is primarily concerned with \emph{intrinsic motivation} -- an experience of interest and enjoyment inherent to an activity -- and the social-contextual factors that support or attenuate its emergence \citep{deci1976notes, deci1985interpersonal, deci1985perceived, deci2000what}. Intrinsic motivation is considered to underpin human development, stimulating exploratory behaviours that identify new modes of interacting with the environment, while reinforcing behaviours that extend these capacities \citep{ryan1997nature}. Past studies have provided evidence of intrinsic motivation's benefits in terms of outcomes such as task performance \citep{ryan1990emotions}, creativity \citep{koestner1984setting}, and persistence \citep{ryan1997intrinsic}. 

CET distinguishes between the satisfactions inherent to an activity (i.e., basic need satisfaction), and \textit{extrinsic rewards}, which are separable from the activity itself (e.g., a cash prize). Extrinsic rewards undermine intrinsic motivation to the extent that their conferral devalues the activity itself and controls further engagement \citep{deci1999meta}. Put another way, extrinsic rewards and other events (e.g., threats) affect the \emph{perceived locus of causality} (PLOC) of the associated activity \citep{decharms1983motivation}: behaviours seen to originate from within the self reflect a more \emph{internal} locus (I-PLOC), whereas behaviours perceived to emanate from outside sources represent a more \emph{external} locus (E-PLOC)\footnote{Note that introjected behaviours are compartmentalised within the self, and are therefore \emph{experienced} as having an E-PLOC despite ostensibly originating from within the individual \citep{decharms1983motivation}.}.

\begin{figure}
\includegraphics[width=0.8\columnwidth]{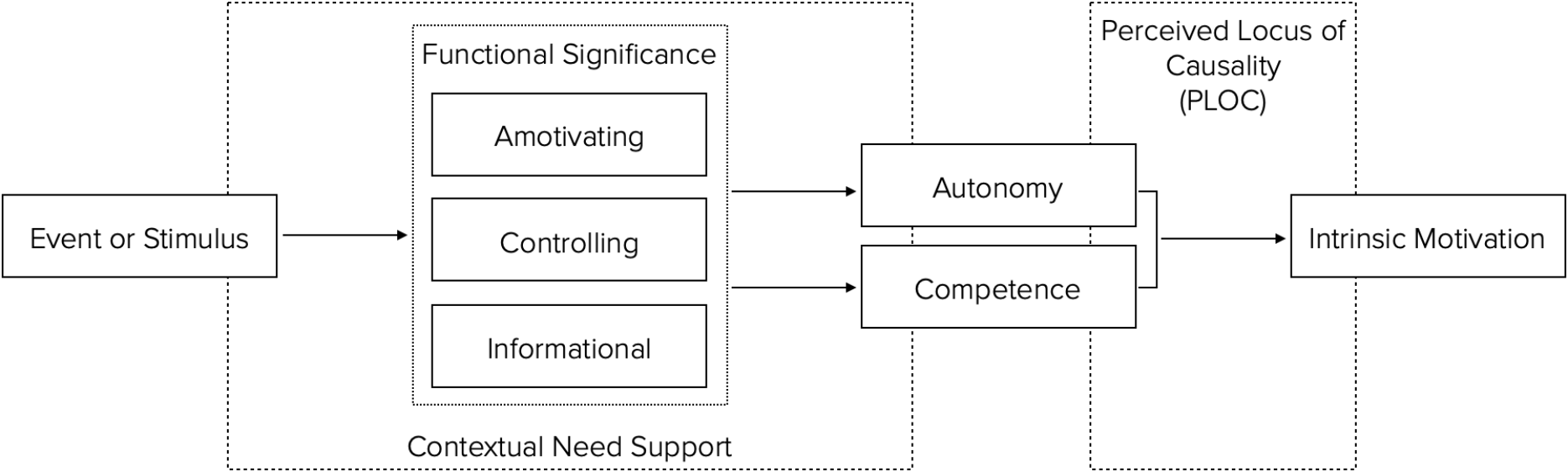}
\caption{The key tenets of cognitive evaluation theory (CET): the extent to which a stimulus influences experienced need satisfaction and intrinsic motivation is determined by its functional significance and the social context. Events perceived as controlling and/or amotivating undermine need satisfaction, promote a more external PLOC, and decrease intrinsic motivation. Events perceived as informational and non-controlling support need satisfaction, promote a more internal PLOC, and increase intrinsic motivation.}~\label{CET_diagram}
\vspace{-2.3em}
\end{figure}

CET further postulates that the effects of rewards and other events depend on the meaning or interpretation the recipient gives to them \cite{ryan2017cet}. That is, each event has a particular \emph{functional significance} for the recipient, defined in terms of how the event impacts experiences of autonomy and competence. For example, a reward could be experienced primarily as a way of \emph{controlling} behavior, in which case it would likely diminish satisfaction of the need for autonomy and undermine intrinsic motivation (see \autoref{CET_diagram}), or it could be experienced as \emph{informational}, affirming or promoting autonomy and 
competence. Some events may instead be \emph{amotivating}, which means the person experiences them as diminishing either a sense of competence, 
autonomy, or both. In particular, rewards reliably undermine intrinsic motivation when their provision is expected; contingent on task engagement, completion, or performance; or when they represent tangible commodities (e.g., money) \citep{deci1999meta}. Intrinsic motivation, conversely, can be improved by small rewards tied to competent behaviour (e.g., coins in the \emph{Super Mario} series) \citep{ryan2017cognitive}, and other forms of feedback likely to be construed as non-controlling and informational \citep[e.g., verbal praise;][]{fong2019negative}.

CET concepts have been typically operationalised via the Intrinsic Motivation Inventory \citep[IMI;][]{centerIMI,mcauley1989psychometric,ryan1982control}, whose subscales include interest/enjoyment, perceived choice, perceived competence, effort/importance, pressure/tension, and value/usefulness. Note that while some SDT scholarship operationalises intrinsic motivation via the interest/enjoyment construct \citep{centerIMI}, it has been suggested that intrinsic motivation is an aggregate function of multiple IMI constructs \citep{mcauley1989psychometric}.

\subsection{Organismic Integration Theory}

\begin{figure*}
\centering
\includegraphics[width=\columnwidth]{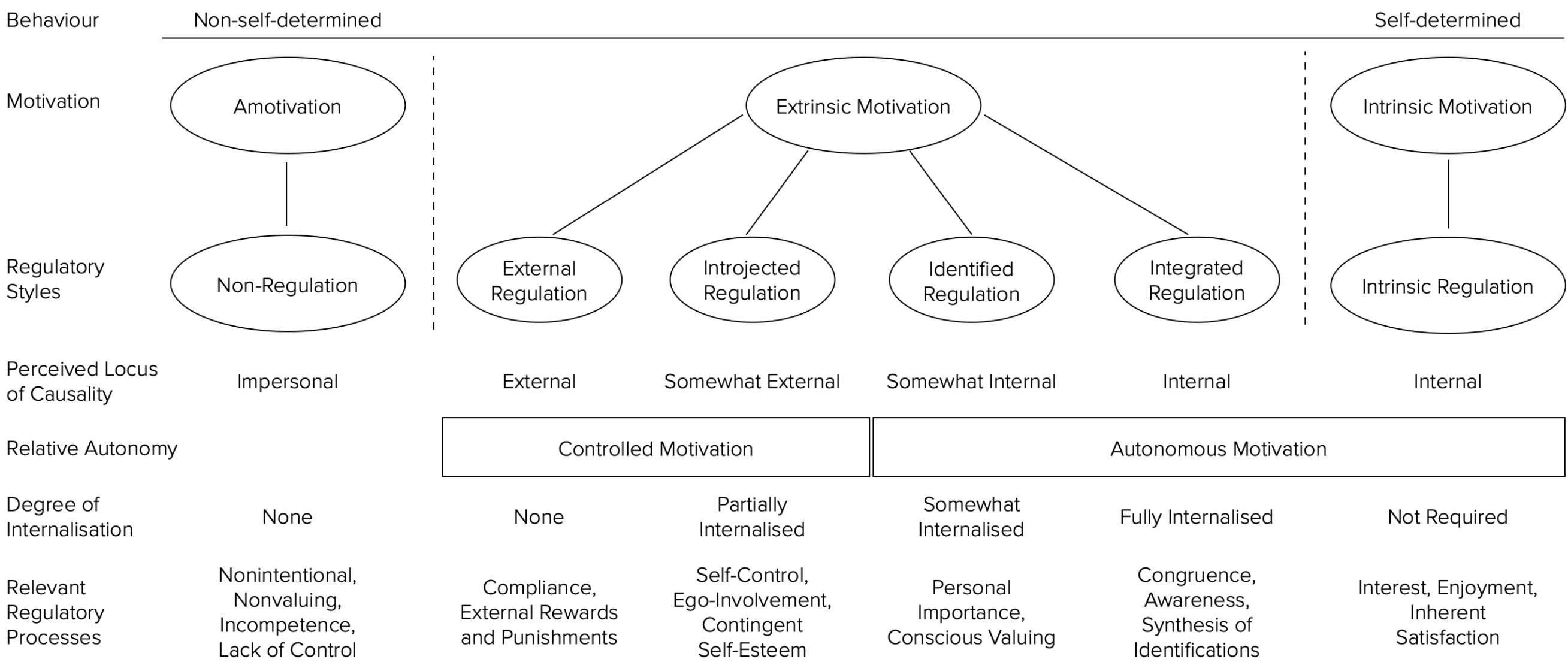}
\vspace{-2em}
\caption{The different motivations and regulations posited by organismic integration theory (OIT), ranging from the least self-determined (amotivation) to the most self-determined (integrated regulation and intrinsic motivation). Adapted from \protect\citep{ryan2017organismic,vansteenkiste2018fostering}.}~\label{OIT_diagram}
\vspace{-2em}
\end{figure*}

Both the \emph{quality} and \emph{quantity} of motivation influence its capacity to energise and direct behaviour \citep{ryan2000intrinsic}. SDT's second mini-theory, \emph{organismic integration theory} (OIT), proposes a continuum model of motivational quality (see \autoref{OIT_diagram}) according to the \emph{relative autonomy} of each form \citep{ryan2017organismic}. OIT attends primarily to the four types of \emph{extrinsic motivation} within this continuum, which collectively pertain to behaviours performed for their instrumental value, and not because they are inherently interesting or enjoyable \citep{ryan2000intrinsic}. Extrinsic motivation is also distinguished from \emph{amotivation}, which pertains to behaviour performed without intentionality.

Extrinsically motivated behaviours are autonomous insofar as they are coherently organised with other facets of the self through \emph{internalisation}, a developmental process facilitated by need satisfaction \citep{deci1994facilitating}. 
\emph{External regulation} is the most controlled form of extrinsic motivation, observed in behaviours ``motivated by and dependent upon external reward or punishment contingencies'' \citep[p.~184]{ryan2017organismic}. \textit{Introjected regulation} obtains in values and behaviours subject to rigid internal controls such as guilt, shame, or contingent self-esteem \citep{deci1994facilitating}. Although introjected regulations are partially internalised, they remain controlling, as their rigidity puts them in conflict with other values and beliefs. 

In contrast, \textit{identified regulation} is a more autonomous form of motivation characterised by the conscious affirmation of behaviours that enact values identified as personally important. Identified regulations may nonetheless remain compartmentalised with respect to other regulations that exist within the self. Identification is hence distinguished from \emph{integrated regulation}, the most autonomous form of extrinsic motivation, which manifests in fully self-congruent and reflectively valued behaviours. Crucially, studies that examine identified and integrated regulations alongside intrinsic motivation \citep{burton2006differential,mankad2014motivational,murray2020mechanisms,sheehan2018associations} suggest that both forms of autonomous motivation predict beneficial psychosocial outcomes \citep[e.g., lower anxiety;][]{sheehan2018associations} and behaviours \citep[e.g., engagement;][]{mankad2014motivational}.

Initially, SDT research measured internalisation by constructing a \emph{Relative Autonomy Index} \citep{grolnick1989parent}, which ``approximates an individual's position along the underlying continuum of self-determination'' \citep[p.~536]{howard2020review} using weighted self-report data. This approach has since been criticised, however, as ``the weights associated with each [motivation] subscale are relatively arbitrary with no published empirical evidence to support them'' \citep[p.~536]{howard2020review}. More granular approaches \citep[e.g.,][]{fernet2020self,howard2018using,howard2021longitudinal} employ 
advanced statistical modelling methods to assess differences in the ways each regulatory style influences behavioural and psychosocial outcomes. Importantly, these latter approaches indicate that the different motivations and regulations posited by OIT are not mutually exclusive. For example, a person can be simultaneously motivated to put effort into their 
work for financial gain and the approval of others, as well as because it aligns well with personal values, and because they find the task interesting \citep{howard2016motivation}. 

The dualistic model of passion \citep[DMP;][]{vallerand2003passions,vallerand2019passion} is sometimes referenced alongside OIT 
\citep{ryan2017identity,vallerand2021reflections}, %
with which it shares 
%
the notion of internalisation. 
According to OIT, extrinsically motivated behaviours become 
autonomous through internalisation (\autoref{OIT_diagram}); 
intrinsic motivation, however, 
does \emph{not} rely on internalisation 
\citep[p.~239]{deci2000what}. DMP instead posits two distinct internalisation processes that pertain 
to behaviours that are liked \citep{vallerand2003passions}. \emph{Harmonious passion} results from autonomous internalisation, where pursuit of a beloved activity is in harmony with aspects of a person's life. \emph{Obsessive passion}, in contrast, follows from controlled internalisation, where enjoyment of an activity is at odds with aspects of the self, and has been linked to compulsive engagement and a range of 
adverse psychosocial outcomes \citep[e.g., 
impaired player wellbeing,][]{johnson2022unsatisfied,przybylski2009having}. Note that although DMP is rooted in SDT \citep{vallerand2021reflections}, 
its co-developers explicitly demarcate the model from OIT
, emphasising that ``irrespective of the type of extrinsic motivation and whether it is internalized in the self (e.g., integrated or identified regulation
), a fundamental difference between passion and extrinsic motivation is the lack of a love for the activity with the latter construct [...] Furthermore, contrary to [DMP], 
no theory or research has hypothesized or found that intrinsic motivation can lead to maladaptive outcomes'' \citep[p.~21]{vallerand2019passion}. 
%


\subsection{Causality Orientations Theory}
SDT is primarily concerned with the ways that individuals are immediately influenced by aspects of their social context. However, 
according to \emph{causality orientations theory} (COT) 
\citep{deci1985general,ryan2017causality,vansteenkiste2010development}, 
people differ in the extent to which they experience their actions in general as self-determined. Three causality orientations are theorised, which co-occur within individuals to varying degrees, yet remain temporally relatively stable. Highly autonomy oriented individuals are more likely to act according to their own interests and values, and interpret events in terms of informational rather than controlling functional significance \citep{deci1985general, vansteenkiste2010development}. Autonomy orientation is therefore positively associated with experiences of competence satisfaction and intrinsic motivation. High control orientation reflects a tendency to act according to external demands, and perceive circumstances as pressuring. Consequently, control orientation is associated with autonomy frustration and less self-determined forms of motivation \citep{ryan2017causality}. Lastly, an impersonal orientation describes a fixation toward obstacles to goal attainment, and a perceived lack of control over outcomes. Outcomes associated with an impersonal orientation include amotivation and ill-health \citep{deci1985general, ryan2017causality}.

A notable extension of COT is the hierarchical model of motivation \citep{blanchard2007reciprocal,vallerand1997toward,vallerand2000deci}, which integrates this global self-determination factor with motivational orientations towards particular life contexts, and individual situations. This line of work theorises each level as exerting a bi-directional effect on its immediate neighbour(s); however, outcome factors (e.g., wellbeing) remain primarily affected by motivation at the same level of generality. In this way, the model relates the individual-difference mini-theories to other research conducted using SDT \citep[p.~233]{ryan2017causality}. 

\subsection{Basic Psychological Needs Theory}
\label{basic}
Autonomy, competence, and relatedness underpin much of SDT's conceptual apparatus. \emph{Basic psychological needs theory} (BPNT) \citep{ryan2002overview} explicates their status as \emph{needs}, and their contributions to growth and wellbeing \citep{vansteenkiste2020basic}. SDT specifies nine criteria that all basic needs must satisfy (see \autoref{tab:bpnt}). These \emph{essential} and \emph{universal} criteria help explain SDT's appeal for scholarship aiming to produce generalisable research findings: need satisfaction should reliably contribute to beneficial organismic processes and wellbeing outcomes, whereas direct proximal threats to basic needs should instead produce dysregulated behaviours and ill-being outcomes with some regularity. \emph{Need frustration}, for instance, denotes the active thwarting of basic needs -- feeling incapable, controlled, or ostracised by others \citep{ryan2017basic},  
and has been shown to predict ill-being outcomes (e.g., burnout, stress) more effectively than need satisfaction \citep{gillet2015effects, haerens2015perceived}.  There is additional predictive value in considering needs as motives \citep{sheldon2011integrating, sheldon2009needs}, whose \emph{directional} qualities orient individuals toward need-supportive social contexts and activities, and away from their need-thwarting equivalents.


SDT propositions on wellbeing are also located in BPNT. The mini-theory locates ``human flourishing'' as an outcome of a well-lived life, that is, through the pursuit of activities that satisfy basic needs \citep{ryan2008living}. This view diverges from other psychological positions -- most notably subjective wellbeing \citep[e.g.,][]{diener2016findings} which conceptualises wellbeing in terms of life satisfaction, positive affect, and low negative affect -- in positing that wellbeing entails a ``focus [...] on self-realisation consistent with the common good'' \citep[p.~59]{ryan2013flourishing}; in particular, ``activity that develops and expresses one’s most reflectively valued and well integrated human potentialities'' \citep[p.~58]{ryan2013flourishing}. For SDT, wellbeing manifests primarily as a sense of aliveness and internal energy \citep[i.e., vitality;][]{ryan1997energy, nix1999revitalization}, and often (somewhat conveniently) subjective feelings of happiness and satisfaction \citep{ryan2017basic}.

Finally, BPNT describes factors that facilitate need satisfaction -- in particular, \emph{autonomy support} and \emph{mindfulness}. Although all need supports are theorised as important for need satisfaction, autonomy-supportive (and controlling) social contexts are considered to operate on all three basic needs \citep{vansteenkiste2020basic}, motivational quality, and other psychosocial outcomes \citep{deci1987support}. Separately, mindfulness is conceptualised in BPNT primarily as a state of open and receptive awareness of the self in interaction with the proximal social context \citep{brown2003benefits}, a state that promotes reflectively valued (i.e., autonomous) behaviour, and hence need satisfaction and wellbeing \citep{rigby2014mindfulness}.

\begin{table}[]
    \centering
    \setlength{\tabcolsep}{2.5pt}
    \renewcommand{\arraystretch}{1.2}
    \scriptsize
    \begin{tabular}{lp{330pt}}
        \emph{Basic criteria} & \emph{Description} \\
        \toprule
         1. Psychological & Basic needs pertain to psychological, rather than physical, qualities of human functioning \\ 
        2. Essential & Need satisfaction contributes to growth, wellbeing, and adjustment; need frustration promotes problem behaviour, ill-being, and psychopathology \\
        3. Inherent & Needs represent evolved aspects of human psychology because their satisfaction confers adaptive advantages \\
        4. Distinct & Needs reflect distinct experiences that emerge independently and irreducibly from the frustration of other needs \\ 
        5. Universal & Need satisfaction and frustration respectively predict well- and ill-being outcomes for \emph{all} people, regardless of demographic factors, personality, cultural background, or need strength \\
        \emph{Associated criteria} & \emph{Description} \\
        \hline
        1. Pervasive & Need-based experiences produce myriad cognitive, affective, and behavioural outcomes that arise at multiple levels of analysis \\
        2. Content-specific & Basic need satisfaction and frustration manifest through specific behaviours and experiences, and are readily expressed in ordinary language \\
        3. Directional & Needs direct and shape human thought, action, and feeling, thereby eliciting the pursuit of need-supportive social contexts while prompting corrective behaviour in need-thwarting situations \\
        4. Explanatory & Basic needs uniquely account for and explain the ways that properties of social contexts influence well- and ill-being outcomes \\
        \bottomrule
    \end{tabular}
    \caption{Qualities of basic needs, as specified in basic psychological needs theory (BPNT). Associated criteria are considered derivative of basic criteria. Material adapted from \citep{vansteenkiste2020basic}.}
    \label{tab:bpnt}
    %
    \vspace{-2em}
\end{table}

\subsection{Goal Contents Theory}
\emph{Goal contents theory} (GCT) pertains to the \emph{types} of outcomes that motivate behaviour, and the ways they influence basic needs, motivation, and wellbeing \citep{ryan2017goal}\footnote{SDT scholars seem to use `goals', `values', and `aspirations' interchangeably.}. 
After causality orientations theory, GCT represents SDT's second individual-difference mini-theory: goal contents can operate at a global level (e.g., amassing wealth), or vary within individual domains (e.g., amassing wealth in \emph{World of Warcraft}). At its core, GCT differentiates between extrinsic goals, which reflect instrumental values (e.g., wealth and fame), and intrinsic goals, which are valuable for their own sake (e.g, personal growth, fulfilling relationships). Intrinsic goals are more strongly linked to beneficial wellbeing outcomes \citep{kasser1996further}; conversely, the pursuit of extrinsic goals is less strongly \citep{martos2014life}, or negatively \citep{kasser2014changes, unanue2014materialism} associated with wellbeing, and negatively related to meaning in life \citep{martos2014life}. These relations between goal contents and psychological outcomes are at least partially mediated by need satisfaction and frustration, respectively \citep{unanue2014materialism, unanue2017when}.

\subsection{Relationship Motivation Theory}
\label{rmt}
\emph{Relationship motivation theory} (RMT) \citep{ryan2017relationships} outlines the qualities that underpin the initiation and maintenance of close relationships. High-quality relationships are characterised by reciprocal need support \citep{deci2006benefits}, autonomous motivation \citep{knee2005self}, and non-contingent care for others \citep{deci2014autonomy}. These relationships satisfy the need for relatedness, and hence improve wellbeing outcomes for those in social contact \citep{lynch2009being}. Conversely, dysfunctional relationships are marked by experiences of need frustration, inconsistent support, or objectification \citep{ryan2017relationships}. Specifically, RMT considers \emph{conditional regard}, whereby relatedness-satisfying behaviours are withheld as a means of control, particularly damaging to relational wellbeing \citep{kanat-maymon2016controlled, roth2012costs}.
Although RMT largely pertains to dyadic interpersonal relations, extensions to group-level processes have also been explored. Initial study findings suggest that basic needs associated with group membership also contribute to individual wellbeing outcomes \citep{kachanoff2020free, kachanoff2019chains}.

\section{SDT Research on Games}
\label{games_sdt}
As part of our expanded view of SDT's applications to games research and design practice, we next look at 
how games have been construed and theorised by \emph{SDT scholars}. 
That is, as well as looking at how HCI scholars apply SDT to games research (see \autoref{Results}), we examine perspectives of scholars \emph{primarily concerned with SDT} who have studied games and who 
arguably introduced SDT to many games researchers \citep[i.e.,][]{ryan2006motivational}. 
We examine these SDT-centric views to identify alternate -- presumably more directed and productive -- uses of theory to those we previously observed at CHI and CHI PLAY \citep{tyack2020self}, and 
identify avenues for theory-informed work 
that have yet been left untapped in HCI games research. 


To our knowledge, the first SDT research involving videogames \citep{arnold1976effects} was conducted shortly after Deci's original formulation of CET \citep{deci1975intrinsic}, and investigated whether extrinsic rewards would reduce intrinsic motivation even for `highly intrinsically motivating' activities such as videogame play. Videogames' intrinsically motivating qualities were also examined in early research on learning \citep[e.g.,][]{malone1981toward}; however, focused examination of other core SDT concepts such as need satisfaction largely began much later \citep{medina2005digital}.

A subsequent run of empirical studies, co-authored by SDT co-developer Richard Ryan, largely aimed to show that SDT's postulates, methods, and concepts could outperform other motivational models \citep{przybylski2009having, ryan2006motivational} and media effects research \citep{przybylski2009motivating, przybylski2014competence} at explaining the widespread appeal of play. These papers typically deployed subscales from the authors' own game-specific measure 
-- the Player Experience of Need Satisfaction scale \citep[PENS;][more below]{rigby2007player} -- and 
the IMI interest/enjoyment subscale \citep{centerIMI}. PENS measures of need satisfaction have been observed to predict game enjoyment, short-term wellbeing, and future play intentions \citep{ryan2006motivational}. This body of work was subsequently summarised in a ``theory-based motivational model'' \citep[p.~154]{przybylski2010motivational}, which Rigby and Ryan also coin the ``PENS model'' \citep{rigby2007player,rigby2011glued,ryan2020motivational}. We next examine the PENS and its components in more detail.


\subsection{Player Experience of Need Satisfaction (PENS)}
\label{pens}

As the name suggests, the PENS model primarily draws from basic psychological needs theory, though initial work also refers to CET \citep{ryan2006motivational}. The key tenet is that games are engaging to the extent that they satisfy players' needs for autonomy, competence, and relatedness. Additionally, the model considers immersion and intuitive controls as defining aspects of the player experience \citep{ryan2006motivational}. The apparent utility of the PENS model is reflected in the high number of citations\footnote{Google Scholar counts over 7000 citations at the time of writing.} that corresponding works \citep[e.g.,][]{przybylski2010motivational,rigby2011glued,ryan2006motivational} have garnered. Moreover, the PENS is featured prominently in materials published by Immersyve \citep[e.g.,][]{rigby2007player}, a UX consulting and market research company co-founded by Rigby and Ryan. 

Unlike other game motivation models \citep[e.g.,][]{bartle1996hearts,yee2006motivations} that are thought to ``largely reflect the structure and content of current games'' \citep[p.~348]{ryan2006motivational}, Ryan and colleagues argue that the PENS is ``agnostic to any specific technology or design'' \citep[p.~172]{ryan2020motivational} in that it accounts for the ``underlying motives and satisfactions that can spark and sustain participation across \emph{all} potential players and game types'' \citep[p.~348, emphasis added]{ryan2006motivational}. 
Further, Rigby and Ryan emphasise the PENS' 
epistemic commitments, in that it
``demonstrates the value of bringing clear psychological theories to game study that can drive real hypothesis testing'' \citep[p.~167]{rigby2011glued}, 
whereas ``not having good theory creates a vacuum, inviting pure speculation about causal connections that might not actually exist
'' \citep[p.~168]{rigby2011glued}. Indeed, the authors strongly imply that hypothesis testing constitutes the most productive application of SDT in games research, arguing that ``research that is not empirically testing specific hypotheses and theory doesn’t do much to advance our knowledge 
 [... of] the relations between gaming and outcomes of interest
'' \citep[p.~167-168]{rigby2011glued}. 

\subsubsection{Need Satisfaction in Games}

Ryan and Rigby posit that ``the impact of nearly every element of game design can be seen as a function of its relations to basic needs
'' \citep[p.~167-168]{ryan2020motivational}. 
The need for competence, for example, is said to be satisfied in scenes of optimal challenge, and by three types of feedback that operate at different temporal scales \citep{rigby2011glued}: \emph{granular} feedback, the immediate feedback players receive for each action, such as blood that appears in response to shooting an enemy; \emph{sustained} feedback, which games use to recognise and reward ``your skill and ability at being consistent'' \citep[p.~24]{rigby2011glued}, such as note streaks in \emph{Guitar Hero} \citep{harmonix2005guitar}; and \emph{cumulative} feedback that ``recognizes the player’s skill and accomplishments in ways that persist'' \citep[p.~29]{rigby2011glued} over the lifetime of the game, such as overall progress. 
Autonomy, in contrast, is said to be satisfied by games that allow players to enact aspects of their \emph{identity} -- or their avatar's -- via self-expression; the degree of perceived choice in \emph{what to pursue} in the game, as in RPG sidequest design; and the degree of \emph{volition} games support through the provision of narrative meaning, structure, and rationales for player behaviour \citep{rigby2011glued}. Finally, games are argued to satisfy relatedness primarily through a sense that the player(-character) \emph{matters} to the game world, 
whereby ``great single-player experiences with NPCs [...] succeed with players partly because they provide thoughtful contingent reactions that successfully yield relatedness satisfactions'' \citep[p.~71]{rigby2011glued}. In multiplayer modes, Rigby and Ryan suggest that relatedness 
is satisfied by mutual support to achieve objectives, or via good-natured, `constructive competition' in which players display sporting conduct. Notably, while these claims about need satisfaction could produce a deeper understanding of how game design can influence player experience, all of the above posited causal relations remain empirically untested in published SDT literature.


\subsubsection{Intuitive Controls} The final two aspects of the PENS model were contrived entirely for games research, and do not otherwise appear in SDT: first, the notion of \emph{intuitive controls} (sometimes `mastery-of-controls' \citep{przybylski2014competence}) refers to the extent to which players feel comfortable learning and becoming proficient at a game's control scheme \citep{ryan2006motivational}. Specifically, intuitive controls have been theorised as  the ``price of admission'' \citep[p.~351]{ryan2006motivational}, that is, a necessary, albeit not sufficient condition for experiencing autonomy, competence and immersion; the relation to relatedness remains unspecified. In practice, the intuitive controls factor has primarily been used to predict \citep[e.g.,][]{ryan2006motivational} and manipulate competence satisfaction \citep{przybylski2014competence}. Moreover, results from regression analyses suggest that the effect of intuitive controls on intrinsic motivation is mediated by competence \citep{przybylski2014competence,ryan2006motivational} and -- occasionally -- autonomy \citep{ryan2006motivational}. Conversely, a lack of intuitive controls has been found to indirectly promote aggressive thoughts and affect by means of impeding competence \citep{przybylski2014competence}. 


\subsubsection{Immersion}
The final PENS concept 
is \emph{immersion} (sometimes `presence' \citep{ryan2006motivational}, or `immersion/presence'; \citep{rigby2007player}), which 
SDT scholars define as ``the illusion of non-mediation'' \citep[p.~350]{ryan2006motivational}, a perspective they draw from prior conceptual work outside SDT \citep[see][]{lombard1997heart}. More specifically, 
SDT offers a tripartite model of immersion that decentres audiovisual quality, and instead corresponds to players' sense of being \emph{physically}, \emph{narratively}, and \emph{emotionally} present in the game world. 
However, only physical presence has a clear basis in prior scholarship \citep[i.e., presence as transportation;][]{lombard1997heart}. The origins of emotional and narrative presence, and their relevance to immersion as defined in SDT games scholarship \citep[i.e., ``the illusion of non-mediation'';][]{lombard1997heart}, 
are not specified. Likewise, immersion's importance for SDT has never been fully explicated -- 
its purported relationships with other theoretical concepts 
remain vague 
\citep[e.g., ``we believe intrinsic motivation in gaming contexts is associated with presence'';][p.~350]{ryan2006motivational}. 
This indeterminacy transfers into how immersion has alternatively been used as a dependent variable, predicted by need satisfaction \citep{przybylski2009motivating,ryan2006motivational}, and as a moderator \citep{przybylski2010motivational,przybylski2012ideal} that purportedly ``magnifies the effects of experiences in virtual contexts on behavior in the real world'' \citep[p.~70]{przybylski2012ideal}. 
We also note that some SDT research \citep{sheldon2015experiential} indicates that mindful awareness -- a key facet of wellbeing in SDT (see section \ref{basic}) -- is inhibited by immersion\footnote{Technically, the study refers to `flow absorption' -- ``being task engaged with minimal self-consciousness and a distorted sense of time'' \citep[][p.~277]{sheldon2015experiential} -- but this definition overlaps considerably with common views of immersion in HCI \citep[e.g.,][]{brown2004grounded,denisova2016convergence}.
}, suggesting that this aspect of SDT games scholarship is actually \emph{inconsistent} with the wider theory. Although SDT scholars have suggested that immersion and mindful awareness do not \emph{inherently} conflict [Rigby, personal communication], the empirical discrepancy is yet to be resolved.

\subsubsection{PENS Scale}
The PENS scale was developed to operationalise players' experience of need satisfaction, immersion and intuitive controls \citep{rigby2007player}, and first formally introduced in \citep{ryan2006motivational}. Competence items reflect how capable players feel during play; autonomy items focus on players' perception of freedom and choice in the game; and relatedness items pertain to the extent to which players feel connected to other players. Due to the scale's proprietary status
, the full list of items has not been made openly available. The most common version appears to be a 21-item scale, where autonomy, competence, relatedness, and intuitive controls are measured with three items each, though some studies employed 5-item versions of the autonomy and competence subscales \citep{ryan2006motivational}. 
Notably, the immersion subscale assesses players' sense of physical, narrative, and emotional presence with three items each \citep{ryan2006motivational}. However, all known instances in SDT games scholarship combine the 9 items into a single mean immersion score \citep{przybylski2009motivating,przybylski2012ideal,ryan2006motivational}. 
Moreover, while it has been claimed that 
the PENS scale was ``
validated in two rounds of confirmatory factor analysis using survey data from 2,000 regular video game players'' \citep[p.~246]{przybylski2009motivating}, to our knowledge, this analysis has never been published. 



\subsection{Other Strands in SDT Games Scholarship}
Outside the PENS model, a somewhat unusual strand of conceptual work led by Rigby \citep{rigby2011glued, rigby2014gamification, rigby2017time} adopts a heroic frame to explain games' motivating and need-satisfying qualities, drawing from Campbell's \textit{Hero with a Thousand Faces} \citep{campbell1947hero}. 
The player as ``learner hero'' \citep{rigby2009virtual} is likewise deployed for game-based learning, as ``a model for more precisely explaining [... virtual worlds'] educational value and motivational appeal'' \citep[p.~221]{rigby2009virtual}. For Rigby, heroic game narratives epitomise 
need satisfaction and intrinsic goal pursuit, 
providing players a contextual frame that ``invites courageous and heroic action [...] communicates that the player’s participation matters
'' \citep[p.~217]{rigby2009virtual}, 
as well as ``opportunity to shine individually [... and] 
help others grow in similar ways'' \citep[p.~219-220]{rigby2009virtual}. Ostensibly, ``such mechanisms [...] 
help quantify the value of the learner hero construct as a touchstone to optimizing self-determination
'' \citep[p.~220]{rigby2009virtual}. However, said `mechanisms' are yet to be empirically examined. 


Another aspect of SDT that uniquely applies to games is the \emph{need density hypothesis}\footnote{To our knowledge, Rigby and Ryan's first articulation of the need density hypothesis dates back to 2011 \citep{rigby2011glued}, but the term \emph{need density hypothesis} made its appearance a few years later \citep{rigby2017time, ryan2017motivation}.}, a claim that game designs are so effectively need-satisfying that they present a risk of overuse to people who experience low need satisfaction in daily life \citep{ryan2017motivation}. In particular, games are considered to present ``just world[s]'' \citep[p.~519]{ryan2017motivation} in that they offer more ``dense, consistent, and immediate'' \citep[p.~529]{ryan2017motivation} experiences of need satisfaction \citep{rigby2011glued, rigby2017time, ryan2017motivation}, relative to other life domains. 
However, SDT scholarship is yet to adequately clarify why videogame play satisfies basic needs more effectively than other activity -- 
evidence supporting the need density hypothesis comes from a small number of interviews, in a larger project that (to our knowledge) has remained unpublished \citep{rigby2011glued}.

A parallel line of empirical work led by Przybylski \citep{przybylski2009having,przybylski2019investigating} likewise investigated the relation between need satisfaction and dysregulated play, linking low levels of need satisfaction in daily life to obsessive passion and impaired psychosocial functioning. While these findings support Rigby's claim that people experiencing low need satisfaction in life are more susceptible to game overuse, they are actually at odds with the need density hypothesis. Evidence \citep{przybylski2019investigating} suggests that players experiencing higher levels of need satisfaction in life derive greater need satisfaction from play. 




\begin{figure}
	\centering
	\includegraphics[width=0.6\columnwidth]{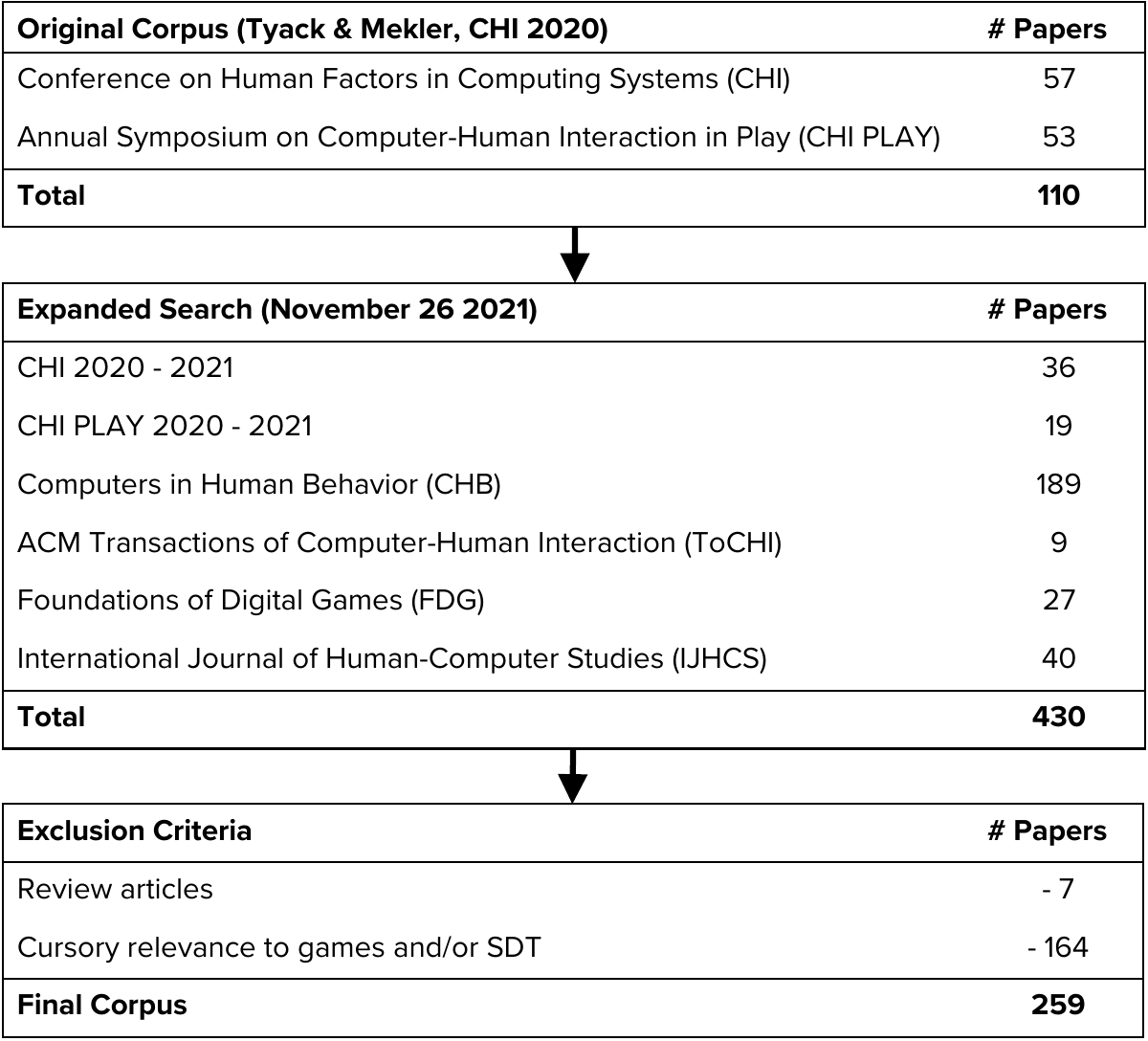}
	\caption{Summary of the literature review procedure.}~\label{review_procedure}
 \vspace{-2em}
\end{figure}

\section{SDT in HCI Games Research}
\label{lit}

This paper takes an expanded view of the ways that SDT has been applied to games -- by HCI games researchers, SDT scholars, and game developers alike -- with an aim to more clearly situate HCI games research within this wider domain. To this end, we review an expanded corpus of HCI games research papers, from a variety of venues. 

\subsection{Literature Review}
\subsubsection{Source selection and search procedure} To update our initial corpus of 110 CHI and CHI PLAY papers \citep{tyack2020self}, we searched the Scopus database and ACM Digital Library on November 26, 2021. We searched Scopus for relevant publications from ACM Transactions of Computer-Human Interaction (ToCHI) and International Journal of Human-Computer Studies (IJHCS), as they are considered high-impact venues in HCI \citep{hornbaek2017technology}. While not unequivocally considered a core HCI publication venue, our search string\footnote{( ALL ( self-determination )  AND  ALL ( theory )  AND  ALL ( game ) )  AND  ( REFAUTH ( ryan )  OR  ( deci ) )  AND  ( LIMIT-TO ( DOCTYPE ,  "ar" )  OR  LIMIT-TO ( DOCTYPE ,  "cp" ) ). This search string is based on the query used in our original review \citep{tyack2020self}.} additionally revealed Computers in Human Behavior (CHB) as the most prolific publisher of SDT-based games research papers among all venues. Many of these papers are in turn cited in HCI games publications, and several highly cited CHB papers pertain to SDT and gamification \citep[e.g.,][]{mekler2017towards, sailer2017how}. We acknowledge that other venues also publish relevant work \citep[e.g.,][]{deterding2015lens,mcewan2020puppy,pusey2022using,turkay2022self}, but decided to focus on what we believe are among the most high-impact venues. 

We additionally searched the ACM Digital Library\footnote{CHI 2021 and CHI PLAY 2021 papers were not yet indexed on Scopus at the time of the search.} for papers from CHI and CHI PLAY published after our earlier literature review \citep[][i.e., proceedings published in 2020 and 2021]{tyack2020self}, as well as 
the Foundations of Digital Games (FDG) proceedings\footnote{Search string: [All: "self-determination"] AND [All: theory] AND [All: game]. Note that the criterion of citing Ryan or Deci was not used for this search, as it was found to wrongly exclude papers \citep[e.g.,][]{kleinman2021gang}.}, a SIGCHI co-sponsored conference covering a wide disciplinary range of games research. FDG proceedings not indexed by the ACM Digital Library were downloaded from each year's conference website. 
Initial filtering of these papers was conducted manually, by the first author, by searching the pdfs for the key terms used when searching Scopus and the ACM Digital Library. Searching in this way returned 
320 new publications that were not found in our previous review, increasing our sample from 110 to 430 papers (see \autoref{review_procedure}).

\subsubsection{Selection of papers for inclusion in the review} Because we were interested in understanding the ways that games research within HCI\footnote{This review takes an inclusive view of what constitutes 'HCI games research' -- a term that here refers to all papers in our corpus -- acknowledging that the 'HCI' status of works at some venues may at times be disputed.} applies and discusses SDT and its concepts, we excluded all review papers, or any papers with only cursory relevance to either SDT or games (n=171, see supplementary material for the full list). Both authors agreed on all exclusions made on this basis. A paper was categorised as cursorily relevant to SDT according to prior citation analysis guidelines; in particular, when it merely ``mentions [some aspect of SDT] in passing, without any context, details, or information'' \citep[p.~10]{girouard2019reality}, or where ``[SDT] work is cited in a list, with no further comment or detail on the individual text'' \citep[p.~831]{marshall2017throwaway}. For example, the work of Gong et al.~\cite{gong2020drives} was excluded for this reason, as it cites SDT only as part of a list of background literature. When assessing relevance to games, we chose to avoid limiting ourselves to a particular definition of ``game'' or ``play'', and accepted any interpretation of ``game-ness'' made by the paper's authors. For instance, we included a paper about an interactive storybook fitness app \cite{saksono2020designing}, on this basis, but excluded another paper focusing on persuasive technology more broadly \cite{villalobos-zuniga2020apps}. We also excluded review papers \citep[such as our own;][]{tyack2020self} to avoid redundancy. As a consequence of these filtering procedures, n=149 papers were retained, 
producing the expanded corpus (N=259) that formed the basis for our subsequent review.

\begin{table}[]
    \centering
    \scriptsize
    \setlength\tabcolsep{3pt}
    \begin{tabular}{l l p{217pt} c}
        Primary & Secondary & Description and Source & Example \\
        \toprule
        Descriptive (
        54.83\%) & Supporting (n=72) & ``[the theory] supports a statement, a simple fact, without necessarily detailing the cited work'' \citep[p.~10]{girouard2019reality} & \citep{schimmenti2017schizotypal} \\
         & Reiterating (n=58) & ``[At least one aspect of SDT] is described . . . The [theory] is presented as valid and reliable and no questions, comments or critique are advanced'' \citep[p.~831]{marshall2017throwaway} & \citep{chen2016influence} \\
         & Methodological (n=76) & ``use of [SDT] materials, equipment, practical techniques, or tools of cited work; use of analysis methods, procedures, and design [based on SDT]'' \citep[p.~66]{bornmann2008citation} & \citep{li2016player} \\
        \midrule
        Analytic (
        37.07\%) & Rationalising (n=88) & ``[the theory] contributes to an argument when it strengthens a line of reasoning, such as to derive a hypothesis, justify a methodological choice, or a data pattern observed'' \citep[p.~11]{girouard2019reality} & \citep{donnermann2021social} \\
         & Contrasting (n=29) & ``citing work contrasts between the current work and [SDT]; citing work contrasts [SDT] with other [theories]; citing work is an alternative to [SDT]'' \citep[p.~66]{bornmann2008citation} & \citep{sweetser2020game} \\
         & Critiquing (n=6) & ``the [theory] is contested . . . in some way engaging or commenting on the [theory] in a way that acknowledges it as something other than uncontested fact'' \citep[p.~831]{marshall2017throwaway} & \citep{burgers2015feedback} \\
        \midrule
        Generative (
        8.11\%) & Extending (n=5) & SDT propositions or concepts are mobilised alongside the 
        citing work to propose theoretical extensions, (re)conceptualisations, or otherwise develop the theory & \citep{koulouris2020effects} \\
         & Designing (n=16) & ``the [theory] inspires or informs . . . design choices'' \citep[p.~12]{girouard2019reality} & \citep{miller2019expertise} \\
        \bottomrule
    \end{tabular}
    \caption{The coding schema applied to the academic corpus (N = 259), largely adapted from prior citation analysis research \citep[i.e.][]{bornmann2008citation,girouard2019reality,marshall2017throwaway}. Note that papers could be coded as belonging to multiple secondary categories.}
    \label{tab:codes}
    \vspace{-2em}
\end{table}

\subsubsection{Coding procedure}
As with our previous review \citep{tyack2020self}, papers were coded with respect to venue, research domain (e.g., gamification), SDT-related measures, theory use (see next paragraph), SDT-related references, SDT concepts named, concept definitions, observed relations involving SDT concepts, statements about SDT, and variations in terminology use. Papers were split between authors, with the first author coding about two thirds of the corpus. Papers co-authored by one of the authors or their (previously affiliated) research groups were coded by the other author, and both independently coded the single paper in the corpus they co-authored \citep[i.e.,][]{tyack2021offpeak}. The coding spreadsheets are included as supplementary material.

In our previous work \citep{tyack2020self}, we broadly coded theory use according to the purposes of HCI theory outlined by Bederson and Shneiderman \citep{bederson2003theories} and Rogers \citep{rogers2012hci}. However, this approach was deemed not sufficiently granular for the aims of the present work, as a theory's intended \emph{purpose} does not necessarily reflect its \emph{application} in the citing work. 
Instead, we adapted the citation typology by Girouard et al.~\citep[]{girouard2019reality}, which was specifically developed for assessing the impact and degree of engagement with theoretical frameworks. Their typology differentiates higher level theory citations (i.e., generative and analytic theory use) -- where theory presumably serves as a direct influence on the citing work -- from low-level citations that draw from theory to help contextualise the work, but where theoretical tenets are otherwise non-critical (i.e., their omission would not substantially change the work). Unlike Girouard et al.~\citep{girouard2019reality}, we did not consider individual mentions of SDT as unit of analysis. 
This would have complicated interpretation due to many papers in our corpus featuring multiple SDT-based references, often in the same sentence. 
Rather, 
each paper in the corpus was assigned a primary theory use category (i.e., descriptive, analytic, or generative, see \autoref{tab:codes}), reflective of its presumed application of SDT. Papers matching several primary theory use categories (e.g., if they featured both descriptive and analytical SDT citations) were assigned the higher level category. 

We further divided 
primary categories of theory use into multiple secondary categories synthesised from prior citation analysis schemata \citep{bornmann2008citation,girouard2019reality,marshall2017throwaway}. For instance, we distinguish `analytic - rationalising' theory use, where SDT concepts and measures are purposefully employed to examine theoretical tenets, from `descriptive - methodological' citations, to account for the prevalence of HCI games papers that employ SDT-based methods without express theory-based rationale \citep{tyack2020self}. 
All 259 papers, including those from our original corpus \citep{tyack2020self}, were (re)coded in this way, with regular meetings to discuss papers and iterate coding categories. Disagreements during coding were minor; all were resolved in agreement. 
For example, we decided to add a `generative - extending' category to account for our interest in work that `talks back' to SDT. 
Given the emphasis on hypothesis testing in SDT games scholarship \citep{rigby2011glued}, papers grouped into the analytic or generative theory use categories were additionally reviewed for SDT-related hypotheses, that is, hypotheses derived from SDT or pertaining to SDT constructs. 


\begin{figure}
	\centering
	\includegraphics[width=0.9\columnwidth]{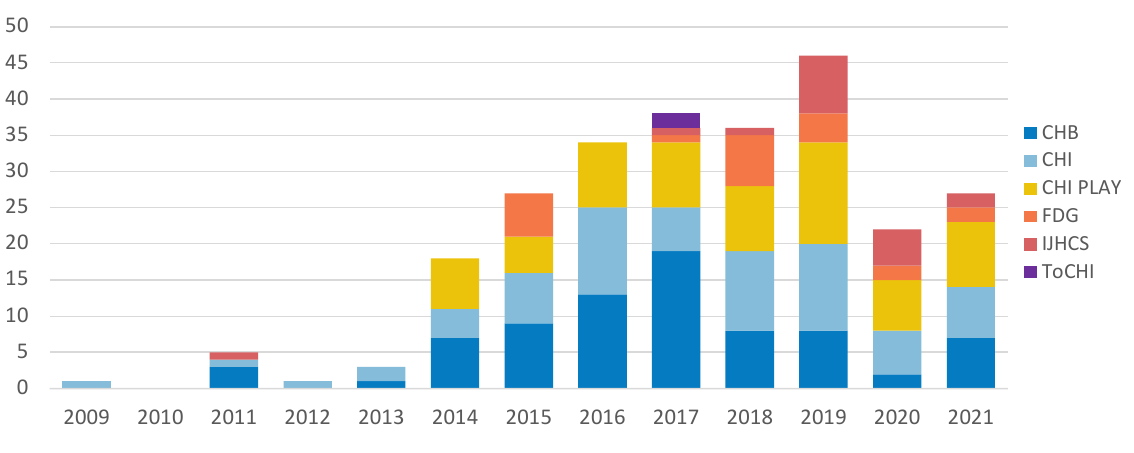}
	\vspace{-1em}
	\caption{Number of reviewed papers (N={259}) referring to SDT per year and publication venue. Note that CHI PLAY proceedings were first published in 2014. One paper published in Computers in Human Behavior in 1993 is not shown here for clarity.}~\label{papers_year}
	\vspace{-2em}
\end{figure}

\section{Results}
\label{Results}
In line with our estimate of the venue's importance, the venue with most papers in the corpus was Computers in Human Behavior (CHB; n=78), followed by CHI (n=70) and CHI PLAY (n=69). Fewer eligible papers were identified from the Foundations of Digital Games conference (FDG; n=22), the International Journal of Human-Computer Studies (IJHCS; n=18), and ToCHI (n=2). Yearly variation in SDT papers published at each venue are shown in \autoref{papers_year}. Overall, the number of papers that cite SDT has increased substantially in the last decade, which we partially attribute to the uptake of the PENS 
questionnaire in HCI games research \citep{tyack2020self}. That said, there is a considerable dip in publications in 2020 and 2021, which we suspect is due to the onset of the global COVID-19 pandemic having substantially impaired researchers' publishing capacity (e.g., due to otherwise increased workload) as well as restricting opportunities for conducting empirical studies on-site. More speculatively, this decrease also coincides with the recently published validation of the Player Experience Inventory \citep{vandenabeele2020development}, which -- while not per se based on SDT -- has been employed as an alternate means to operationalise competence and autonomy need satisfaction in games \citep[e.g.,][]{hicks2019juicy,tyack2021small}, and exhibits more consistently solid psychometric properties than the PENS \citep{bruhlmann2015how,johnson2018validation}.

Although we observed a fairly wide variety of application domains, almost half (n=125; 48.26\%) of all papers in our corpus pertained to player experience. 
Other popular topics included gamification (n=49; 18.92\%), design (n=39; 15.06\%), VR/AR (n=28; 10.81\%) and game-based learning (n=26; 10.04\%).

In the following, we report our analysis of the 259 papers reviewed, focusing on which SDT concepts were mentioned
, as well as how the theory has been applied. Where relevant to theory use, we also comment on SDT-based measures and theoretical misunderstandings, but refrain from a comprehensive description, as findings largely resemble those from our previous review \citep{tyack2020self}. 

\subsection{Prevalence of SDT Concepts and Measures}
\label{prevalence}

\begin{figure}
	\centering
	\includegraphics[width=1.0\columnwidth]{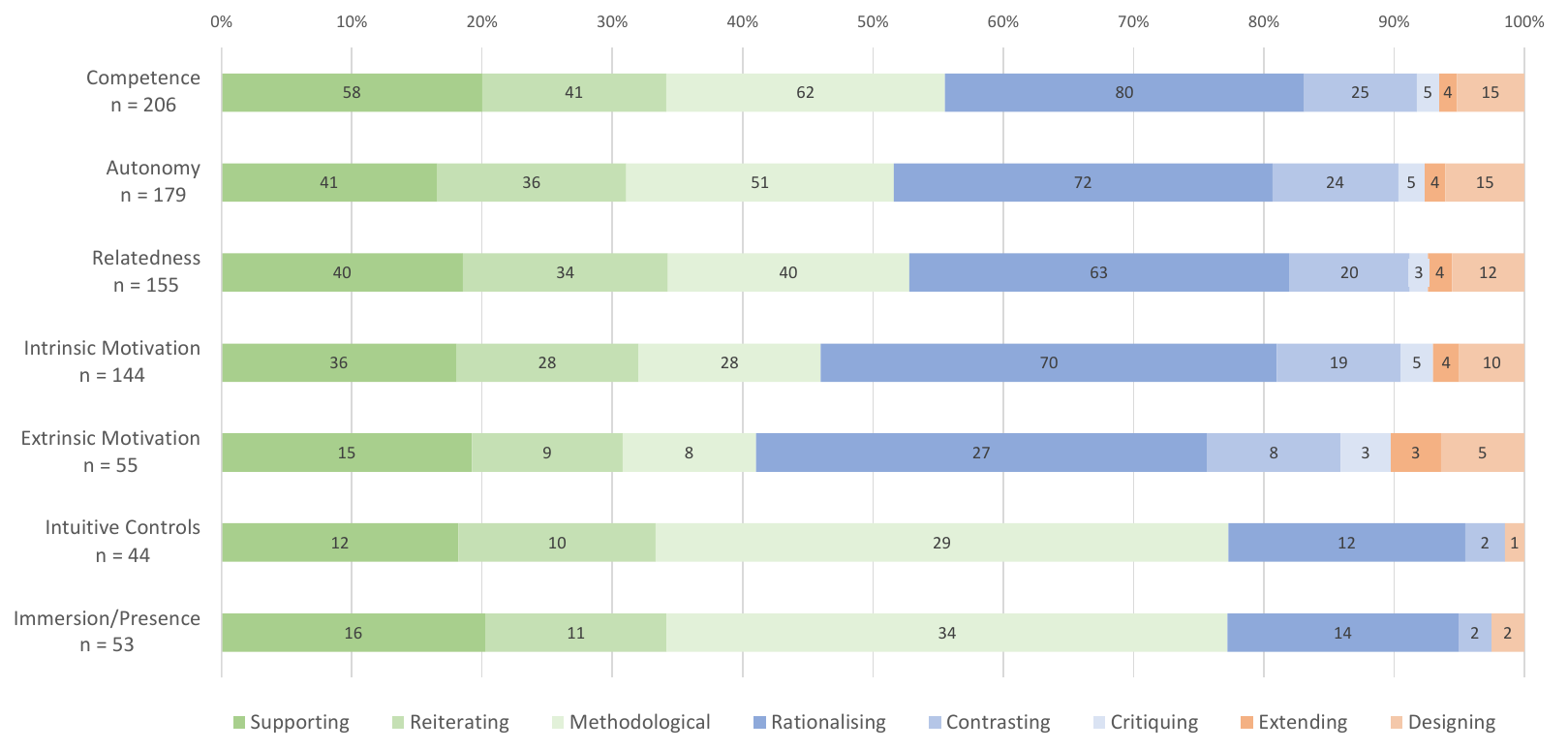}
	\vspace{-2em}
	\caption{The seven most frequently \emph{mentioned} SDT concepts in the HCI games corpus, distributed per secondary theory use category. Note that this does not readily correspond to the concept having been \emph{used} in that manner \citep[e.g., we coded][as an example of rationalising theory use, as it included hypotheses about autonomy. The study, however, also measured relatedness without express theoretical rationale]{smeddinck2016difficulty}. Numbers do not add up to 100\%, as papers could include multiple instances of secondary theory use.}~\label{theory_concept}
	\vspace{-2em}
\end{figure}

Need satisfaction and intrinsic motivation featured prominently in the expanded corpus (
\autoref{theory_concept}) -- primarily competence (n=206; 79.54\%), followed by autonomy (n=179; 69.11\%), relatedness (n=155; 59.85\%), and intrinsic motivation (n=144; 55.60\%). 
Concepts related to extrinsic motivation (n=55; 21.24\%), such as internalisation (n=17; 6.56\%) and its associated regulatory styles were less commonly present. As observed in our earlier review \citep{tyack2020self}, SDT mini-theories were rarely mentioned, despite many papers' application of their foundational concepts. 
Cognitive evaluation theory received most mentions (n=13; 5.02\%), followed by organismic integration theory (n=7; 2.70\%), and causality orientations theory (n=1; 0.39\%). Other SDT concepts mentioned include need frustration \citep[n=4; 1.54\%, i.e.,][]{allen2018satisfaction,mills2020self,tyack2020restorative,tyack2021small}, PLOC \citep[n=2; 0.77\%, i.e.,][]{cruz2017need,wang2011understanding}, and functional significance \citep[n=1; 0.39\%, i.e.,][]{vanroy2019collecting}. Goal contents theory was referenced indirectly once by mention of extrinsic goals 
\citep[i.e.,][]{kajastila2016augmented}. No mention was made of relationship motivation theory.

With regards to concepts specific to SDT games scholarship, immersion/presence (n=53; 20.46\%) and intuitive controls (n=44; 16.99\%) received relatively few mentions, despite their status as key components in the PENS model. The need density hypothesis was cited twice \citep[0.77\%, i.e.,][]{allen2018satisfaction,tyack2021small}; no reference was made to the hero construct. 

Overall, 54.05\% (n={140}) of the reviewed papers employed at least one measuring instrument based on or adapted from SDT. The IMI was used in 31.66\% (n=82) of the reviewed papers, with interest/enjoyment (n=76; 29.34\%) the most commonly measured construct (
\autoref{tab:measures}). The PENS\footnote{To our knowledge, at least three versions of the PENS have seen use in HCI games research -- the PENS v1.6 \citep[e.g.,][p.~43]{johnson2018validation}, whose commercial copyright is dated 2007; a slight variation found in a 2011 doctoral thesis \citep[p.~101]{przybylski2011dispositional}; and one with 4-item competence and autonomy subscales \citep[p.~201]{neys2014exploring}, apparently provided by the scale owners \citep[p.~208]{neys2014exploring}. The scale's commercial copyright, and inadequate citation practices, complicate identification of the PENS version used in most papers.} was employed in 25.87\% (n=67) of papers reviewed, with competence 
and autonomy (each n=62, 23.94\%) dimensions most frequently assessed. While the PENS and IMI were by far the most popular SDT-based measures used, 
many papers assessed SDT concepts by alternative means. Examples include the Ubisoft Perceived Experience Questionnaire (UPEQ) \cite{azadvar2018upeq,evin2020enabling}, Player Experience Inventory (PXI) \cite{hicks2019juicy,tyack2021small}, and Basic Psychological Need Satisfaction scale (BPNS) \cite{Depping2018friendship,goh2017perceptions}; intrinsic motivation via measures of free-choice behaviour \cite{Birk2016fostering,Bowey2017stories}, and scale items adapted from other measures \cite{Bian2018weak,Palacin-Silva2018gamification}; intrinsic and extrinsic motivations via the Situational Motivation Scale (SIMS) \cite{alexandrovsky2021serious,rodrigues2021personalisation} and Gaming Motivation Scale (GAMS) \cite{mills2020self,Shaer2017gaming}.






\subsection{Descriptive Theory Use}

Descriptive (n=142, 54.83\%) papers, which made relatively limited use of SDT, comprised a sizeable part of our corpus. These papers are not of primary interest to the present work, as they largely resemble those 
detailed in our previous review \citep{tyack2020self}: Papers coded as descriptive mostly deployed SDT to contextualise their work in terms of psychological theory, \emph{describing} core tenets but without properly explicating or positioning them in the context of the work. 

In an example of \emph{supporting} theory use, Schimmenti et al. cite SDT only to highlight that ``many studies have explored the motivations of [MMORPG] gamers'' \citep[p.~286]{schimmenti2017schizotypal}, and that ``motivations for playing [...] always emerge as a result of psychological and contextual factors'' \citep[p.~288]{schimmenti2017schizotypal}. Similarly, in a section outlining the varied appeals of videogame play, Chen and Sun cite Ryan et al.~\citep{ryan2006motivational} to \emph{reiterate} a common SDT claim that ``providing choice within a game has the potential to enhance a player's perception of autonomy, which has been shown to increase intrinsic motivation'' \citep[p.~342]{chen2016influence}. This statement, however, is not revisited in the remainder of the paper -- nor does the cited SDT paper \citep{ryan2006motivational} actually demonstrate the impact of choice on autonomy or intrinsic motivation. Notably, while several of these works dedicated considerable space to describing theoretical concepts \citep[e.g.,][]{kappen2017gamification,passmore2020cheating}, omitting SDT would not have substantially changed the research.

As in our prior corpus, \emph{methodological} uses of SDT -- where SDT-based measures were employed without further consideration of theoretical tenets -- were also a strong incentive for citation. Most mentions of intuitive controls and immersion were in this category (see \autoref{theory_concept} and \autoref{tab:measures}), in reference to the PENS subscales \citep[e.g.,][]{kao2021evaluating,nagle2021pathfinder,weech2020narrative}. 
Kao et al., for instance, resorted to the PENS and IMI scales to evaluate a series of VR game tutorials, as ``
[PENS] and intrinsic motivation are both well-established frameworks with extensive empirical validation'' \citep[p.~12]{kao2021evaluating}. However, no explanation is provided as to why need satisfaction or intrinsic motivation might improve in response to different design elements, nor how intuitive controls or immersion relate to this.

A notable case of descriptive theory use is the validation of the Ubisoft Perceived Experience Questionnaire \citep[UPEQ;][]{azadvar2018upeq}, which has been presented as an alternative to the PENS need satisfaction subscales. Unlike the PENS, the UPEQ additionally captures players' sense of relatedness to game characters. Azadvar and Canossa further posit that autonomy, competence, and relatedness constitute ``correlates of sustainable positive interaction with the game [that] benefit game developers in offering them feedback that is not only game-oriented and actionable but also does not hinder the creative process of game development'' \citep[p.~3]{azadvar2018upeq}. 
However, while the UPEQ was found to explain a small amount of variance for several behavioural outcomes (e.g., number of days played), 
the work does not engage with SDT tenets to theorise about why need satisaction might be predictive of said outcomes. 

Similar to our prior findings \citep{tyack2020self}, we observed some unusual interpretations of SDT in the descriptive category, which pertained to research methods and misunderstandings of SDT terminology. One paper, for example, measured need satisfaction via an unknown combination of items seemingly derived from the PENS and BPNS self-report measures \citep{pe-than2014making}. Another paper (mis)characterised basic needs as ``types of motivation'' \citep[p.~1210]{chen2016scaffolding} related to learning, but this did not figure in the study otherwise, which relied on the IMI to measure competence, autonomy, and interest/enjoyment. 

In short, descriptive theory use was common in our corpus, but divorced from deeper engagement with SDT and largely inconsequential to the actual research. Instead, authors cited theoretical concepts to broadly contextualise their own work, or deployed SDT-based measures as purportedly well-established player experience evaluation methods.

 \begin{table}[]
 \centering
\scriptsize
 \setlength\tabcolsep{3.3pt}
\begin{tabular}{clrrlclrrlclrr}
\toprule
\multicolumn{4}{c}{Descriptive (n=142)} &  & \multicolumn{4}{c}{Analytic (n=96)} &  & \multicolumn{4}{c}{Generative (n=21)} \\
\cmidrule{1-4} \cmidrule{6-9} \cmidrule{11-14}
\multicolumn{1}{l}{Scale} & Subscale & N & \% &  & \multicolumn{1}{l}{Scale} & Subscale & N & \% &  & \multicolumn{1}{l}{Scale} & Subscale & N & \% \\
 \cmidrule{1-4} \cmidrule{6-9} \cmidrule{11-14}
\multirow{6}{*}{\begin{tabular}[c]{@{}c@{}}IMI\\ (n=45)\end{tabular}} & interest/enjoyment & 44 & 16.99 &  & \multirow{6}{*}{\begin{tabular}[c]{@{}c@{}}IMI \\ (n=36)\end{tabular}} & interest/enjoyment & 31 & 11.97 &  & \multirow{6}{*}{\begin{tabular}[c]{@{}c@{}}IMI \\ (n=1)\end{tabular}} & interest/enjoyment & 1 & 0.39 \\
 & effort/importance &  {24} & {9.27}  & &  & effort/importance &  {15} & {5.79}  &  &  & effort/importance & {1} & {0.39}  \\
 & pressure/tension &  {24} & {9.27} &  &  & pressure/tension &  {15} & {5.79}  &  &  & pressure/tension &  {1} & {0.39}  \\
 & value/usefulness &  {6} & {2.32} &  &  & value/usefulness &  {2} & {0.77}  &  &  & value/usefulness &  {0} & {0.00}  \\
 & perceived competence &  {18} & {6.95}  &  &  & perceived competence &  {19} & {7.34}  &  &  & perceived competence &  {1} & {0.39}  \\
 & perceived choice &  {6} & {2.32}  &  &  & perceived choice & {7} & {2.70} &  &  & perceived choice &  {0} & {0.00}  \\
 \cmidrule{1-4} \cmidrule{6-9} \cmidrule{11-14}
\multirow{5}{*}{\begin{tabular}[c]{@{}c@{}}PENS \\ (n=41)\end{tabular}} & competence &  39 & 15.06 &  & \multirow{5}{*}{\begin{tabular}[c]{@{}c@{}}PENS \\ (n=25)\end{tabular}} & competence &  {22} & {8.49}  &  & \multirow{5}{*}{\begin{tabular}[c]{@{}c@{}}PENS \\ (n=1)\end{tabular}} & competence &   {1} & {0.39}  \\
 & autonomy &  {37} & {14.29}  &  &  & autonomy &  {24} & {9.27} &  &  & autonomy &  {1} & {0.39} \\
 & relatedness &  {27} & {10.42}  &  &  & relatedness &  {18} & {6.95} &  &  & relatedness &  {1} & {0.39} \\
 & presence/immersion &  {26} & {10.04}  &  &  & presence/immersion & {10} & {3.86} &  &  & presence/immersion &  {1} & {0.39} \\
 & intuitive controls &  {26} & {10.04} &  &  & intuitive controls & {9} & {3.47} &  &  & intuitive controls & {1} & {0.39} \\
 \bottomrule
\end{tabular}
\caption{IMI and PENS use in the HCI games corpus, split by primary theory use category. Note that for the analytic and generative categories this does not readily correspond to the measure having been \emph{applied} in that manner \citep[e.g., we coded][as an example of generative theory use, but the IMI measure was deployed for analytic - rationalising purposes]{koulouris2020effects}. Data do not sum to 100\% (N={259}), as papers did not include all subscales or used multiple measures.
}~\label{tab:measures}
\vspace{-3em}
\end{table}

\subsection{Analytic Theory Use}
This category refers to instances where SDT tenets were purposively tied to the citing authors' work, often to inform the research in some way. As such, it shares much overlap with the ``explanatory'' theory use category described in our previous review \citep{tyack2020self}. Overall, analytic uses of SDT were relatively well-represented (n=96, 37.07\%), particularly in the expanded corpus, 
with over 40\% of instances found in CHB publications.

\subsubsection{Rationalising}
While the bulk of papers in our corpus made limited use of theory, a fair amount of works (n = 88) explicitly deployed SDT as the cornerstone of the research, with literature reviews that tended to emphasise relevant theoretical tenets and concepts, which informed hypotheses (n=44) or research questions. These could be relatively straightforward, as in Kim et al., who predicted that (among other things) ``feelings of autonomy will be positively related to enjoyment'' \citep[p.~696]{kim2015sense}. More elaborate hypothesising is found in work by Smeddinck et al.~\citep[p.~5597]{smeddinck2016difficulty}. After introducing some key tenets of SDT and flow theory, the authors posited that the provision of embedded difficulty choices would increase players' sense of autonomy and immersion, compared to conditions where gameplay was interrupted by non-diegetic game elements (i.e., a menu prompting difficulty choice). In alignment with the study's theoretical grounding, the dependent variables were operationalised via the PENS autonomy and immersion subscales. 


Other works linked their hypotheses explicitly to SDT mini-theories \citep[e.g.,][]{attig2019track,burgers2015feedback,mills2020self,neys2014exploring}, for instance, drawing from CET and COT to theorise about the extent to which gamification elements affect competence: ``Given the assumption that points, levels and leaderboards afford competence need satisfaction [...] the effect should be more pronounced for levels and leaderboards, since they provide more performance feedback than points only [... further,] 
causality orientation may moderate the effects of feedback
'' \citep[p.~528]{mekler2017towards}. Beyond the derivation of hypotheses, SDT was also 
used 
to post-hoc justify results consistent with its theoretical tenets \citep[e.g.,][]{goh2017perceptions,groening2019achievement,li2019book}. For example, in an OIT-informed study of pro-environmental behaviour, Van Houdt et al. \citep[p.~250]{vanhoudt2020disambiguating} conclude that their study ``lends support to the hierarchical model of SDT [...] 
that someone’s personality or user type, as for example defined by the Hexad, [...] 
may not be sufficient to predict gamification preferences''. 






A few works resorted to SDT to inform their research methodology. Unlike papers in the descriptive - methodological category, these works purposively applied the theory to ``
understand(ing) day-to-day play experiences in terms of existing SDT concepts'' \citep[p.~8]{tyack2021small}. 
Saksono et al., for example, employed relatedness need satisfaction as a lens for interpreting interview data around the use of a playful family fitness app \citep{saksono2020designing}. Tyack and Wyeth, investigating the role of autonomy need satisfaction for short-term well-being \citep{tyack2021small}, 
detailed how their experimental manipulation was tailored according to SDT tenets 
to deliberately thwart autonomy: 
%
``As SDT indicates that ``control [...] can be done to people by contextual events and is therefore more easily evidenced [than autonomy support]” \citep[p.~1027]{deci1987support}, the design process largely focused on 
[...] the presence or absence of pressuring directives
, pre-determined puzzle solutions, and elements that limit choice over in-game behaviour'' 
\citep[p.~7]{tyack2021small}. In another example, Tyack 
et al.~outlined how SDT informed selection of a game intended to 
``
consistently support competence regardless of player skill, while providing fewer opportunities to satisfy other basic needs. Previous SDT research \citep{ryan2017motivation,ryan2006motivational} identified game controls, progression structure, and genre as particularly relevant'' \citep[p.~4]{tyack2020restorative}.

Although papers in the rationalising category generally applied SDT in substantial ways, they also frequently contained unorthodox theoretical claims and methodological practices. Luu and Narayan, for example, critiqued a ``meta-analysis [that] included studies in which broad measures of motivation, interest, engagement, and/or even attitudes towards tasks were used as measures for motivation in serious games'' \citep[p.~111-112]{luu2017games}. In their own study, however, they combined all 21 items of the PENS scale (i.e., which index need satisfaction, immersion, and intuitive controls -- not intrinsic motivation) to assess intrinsic motivation, engaging in the same questionable research practice. 

SDT concepts were also (mis)characterised in somewhat unusual ways. In one instance, a study of young women's play habits and undergraduate degree selection \citep{hosein2019girls} includes a number of 
spurious connections between basic needs and other variables, such as self-concept (competence), income deprivation (autonomy), and time spent playing videogames (relatedness). Similar conceptual flexibility is demonstrated in a case study of frustration in `casual games' \citep{roest2015engaging}, which links 
competence to the illusion of control, and 
suggests that relatedness is satisfied by competition, ``even though players cannot really compete, the illusion of control bias lets them believe they can'' \citep[p.~5]{roest2015engaging}. 

Finally, the presence of theory-based hypotheses in a paper did not preclude descriptive use of \emph{some} SDT concepts. For instance, although the aforementioned study by Smeddinck et al.~\citep{smeddinck2016difficulty} is exemplary for how transparently it grounds its hypotheses in SDT, the same study also included PENS measures for competence, relatedness, intuitive controls, as well as the IMI interest-enjoyment and effort-importance subscales, with no apparent theoretical rationale other than ``as additional dimensions rooted in SDT to augment the PENS results'' \citep[p.~5599]{smeddinck2016difficulty}.


\subsubsection{Contrasting}
\label{contrast}
Despite the fair number of papers making use of SDT to inform their studies and explain their data patterns, we observed comparatively little interest (n = 29) in 
discussing findings in terms of their implications \emph{for} SDT, beyond a common acknowledgement that results were (largely) consistent with the theory. Indeed, across theory use categories, results that did \emph{not} align with SDT were consistently blamed on other factors -- study design \citep[e.g.,][]{sailer2017how}, quirks in the data \citep[e.g.,][]{poppelaars2018impact}, 
or by recourse to other (sometimes previously-unmentioned) work \citep[e.g.,][]{dunbar2018reliable}. 
Mekler et al.~\citep{mekler2017towards}, for example, tested SDT-derived hypotheses that gamification would increase competence and intrinsic motivation in an image annotation task -- for which they did not find support. They follow this up with a lengthy discussion of SDT-conforming explanations under which their hypotheses might nevertheless hold true, e.g., ``the motivational appeal of many games lies in their ability to provide players with challenges to master, hence allowing them to experience feelings of competence. The image annotation task, on the other hand, could hardly be considered challenging'' \citep[p.~532]{mekler2017towards}.

Of course, it is informative to first consider methodological confounds and how divergent results might be explained within SDT. What is striking about the aforementioned examples, however, is the apparent 
reluctance to 
consider explanations that may point to potential gaps or incongruities in the theory -- and could drive further theory development and specification. 
Correspondingly, we observed only few instances that more directly contrasted their findings with SDT. Tyack and Wyeth \citep{tyack2021small}, for example, discuss how ``the observed variation in emotional trajectories (H2) speaks to the difficulty of characterising these experiences as unequivocally positive or negative [...] this interpretation is complicated by their respective associations with need satisfaction and frustration (H3) – which are identified as unequivocally positive and negative in SDT'' \citep[p.~18]{tyack2021small}. 


Finally, some papers placed SDT in conversation with other theories 
\citep[e.g.,][]{burgers2015feedback,mills2020self}. Liao et al.~\citep{liao2020impacts}, for example, fruitfully combined aspects of SDT with self-affirmation theory to examine links between real-world need satisfaction, self-esteem, and future play intention. Other papers in this category discuss multiple theories to make wider conceptual points, as in Sweetser and Aitchison's work on (extrinsic) reward-based systems and intrinsic motivation \citep{sweetser2020game}, which compares perspectives from behaviourism, SDT, and flow theory. However, it is not always clear why authors have decided to combine SDT with other theories. A notable exception is work by Poeller et al.~that combines SDT and Motive Disposition Theory \citep{poeller2018implicit,poeller2021seek}, ``because SDT [...] 
does not focus on individual need strength. Therefore, we give an example on how different motivational theories have greater explanatory power when used together
'' \citep[p.~14]{poeller2021seek}. 

\subsubsection{Critiquing}
Finally, we observed few papers in our corpus (n = 6) that contested aspects of SDT, and even these tended to be rather indirect. For example, investigating gamified achievements, Groening and Binnewies note that they ``do not want to get into the controversial debate if tangible rewards undermine motivation'', but regardless ``want to emphasize that we believe digital achievements may be more than just goals with rewards'' \citep[p.~162]{groening2019achievement}. Our own work \citep{tyack2021offpeak} 
is also relatively 
circumspect, in that it largely 
serves to critique SDT’s exceptionalist characterisation of videogames by drawing attention to Rigby and Ryan's claim \citep[p.~109]{rigby2011glued} that 
play can be ``comparable to recreational drug use'' \citep[p.~2]{tyack2021offpeak}. 

However, work by Palomäki et al.~is notable for its more explicit disagreement with SDT; specifically, the authors interpret their study results as complicating the relationship between competence and intrinsic motivation, arguing that ``as one gains experience in an activity,  
expectations rise accordingly 
[... which] will most strongly determine the (dis)satisfaction one derives from the task, rather than one’s absolute level of perceived competence'' \citep[p.~10]{palomaki2021link}. Other instances of more express critique include questioning the ``increasingly prescriptive links between need satisfaction and specific game elements'' made in SDT games scholarship \citep[p.~19]{tyack2021small}, or challenging the explanatory capacity of SDT, e.g.: 
``
the need for relatedness is a universal need that all individuals share. Yet, SDT -- or its myriad mini-theories -- provides little explanation of how people prefer to satisfy their universal need to relate to others'' \citep[p.~2]{poeller2021seek}. 







\subsection{Generative Theory Use}
We coded papers as generative when SDT was leveraged to engender new conceptual insights or inform design. As in our previous review \citep{tyack2020self}, generative papers
formed only a fraction (n=21, 8.11\%) of those reviewed. 

\subsubsection{Extending Theory} 
Only five recorded efforts aimed to \emph{extend} SDT in some way \citep[i.e.,][]{deterding2016contextual,dindar2014motivational,koulouris2020effects,orji2018personalizing,tondello2016gamification}. Dindar and Akbulut \citep{dindar2014motivational}, for example, attempted to map SDT's basic needs onto Yee's motivational framework \citep{yee2006demographics}, arguing that ``even though these motivational needs are summarised in previous works [...] interrelationships among these motivations in video-gaming environments have not been described yet'', and that their study results ``can guide further empirical works to scrutinise the interrelationships among these basic needs'' \citep[p.~124]{dindar2014motivational}. Koulouris et al.~\citep{koulouris2020effects} studied avatar identification in a VR exergame, deploying SDT as one of many theoretical lenses through which to explain the study results. In doing so, the authors also draw from other exergame concepts as well as Jungian psychology to propose two new SDT-related terms: ``\emph{intrinsic identiﬁcation}, which is based on a recognition of one’s true Self and fostering a sense of oneness with the Self, and \emph{extrinsic identiﬁcation}, which is based on personas projected by the ego and drawing the attention away from the Self'' \citep[p.~10; emphasis in original]{koulouris2020effects}.

Although these papers show sufficient investment in SDT to propose theoretical extensions -- or perhaps as a consequence of doing so -- both works contain claims about autonomy that seem to bear little relation to its characterisation in SDT. When linking basic needs to Yee's framework, Dindar and Akbulut argue \citep[citing Rigby and Ryan,
][]{rigby2011glued} that because ``video-game players can form an idealized view of their personalities and have novel experiences which are somehow impossible to find in their real lives'', that ``the immersion component of Yee’s framework can be considered equivalent [to] the autonomy need'' \citep[p.~121]{dindar2014motivational}. However, the cited reference does not actually discuss\footnote{Rigby and Ryan emphasise the importance of `authenticity' for immersion \citep[pp.~84-88]{rigby2011glued}, where authenticity refers to the game world acting and reacting to the player in believable and consistent ways that afford `genuine' need satisfaction. Perhaps Dindar and Akbulut meant to  cite Przybylski et al.~\citep{przybylski2012ideal} instead, who found that immersion moderated the relation between intrinsic motivation and ideal-self---game-self convergence, i.e., when players' in-game self concept reflects aspects of their ideal self in life. However, this work did not examine the role of autonomy.} the notion of players' ideal self, nor specifically link immersion to autonomy need satisfaction. Koulouris et al. instead refer to SDT when discussing their study results, suggesting that ``realistic avatars better fulfill the three main psychological needs that facilitate self-motivation'', including ``autonomy because a realistic avatar frames their efforts as self-competition independent of others'' \citep[p.~9]{koulouris2020effects} -- the theoretical basis of which is left unexplained. In contrast, Deterding's grounded theory of contextual autonomy support in play \citep{deterding2016contextual} hews more closely to 
CET, outlining testable propositions of how the presence or absence of situational demands (e.g., 
making time for play) impacts autonomy need satisfaction. 


\subsubsection{Designing}
\label{Designing}
Several authors used SDT as a tool for \emph{design} (n=16, 6.20\%) -- both as part of the design process itself, and (usually following a user study) to formulate design implications. 
The ways that SDT actually \emph{informed} design in these works was not always entirely clear, however, as in Fatehi et al., who ``were inspired in our design approach by considering \emph{player motivations}'' \citep[p.~3; emphasis in original]{fatehi2019gamifying} -- which here refers to an amalgam of Bartle's player typology \citep{bartle1996hearts}, the Hexad model \citep{tondello2016gamification}, and SDT. The authors' actual design description makes only passing mention of (an unspecified type of) motivation; for example, in noting that they ``tried to characterise the writer and the critic [... to] provide a narrative context that would motivate [participants] for the task'' \citep[p.~5]{fatehi2019gamifying}.



Even when artefact designs were explicitly identified as being `informed' by SDT, however, these connections were often limited to the inclusion of particular game features (e.g., points, badges, and leaderboards). In their study of gamified learning systems, for example, van Roy and Zaman note that ``by awarding group points we also intended to support students’ feelings of competence, as these points served as reinforcing feedback'' \citep[p.~41]{vanroy2019unravelling}. 
The relative terseness of these theoretical links often stood in contrast with more thorough descriptions of the designs as a whole, typically supplemented with arrays of annotated screenshots and (in some cases) a link to the artefact itself. 

That said, some works stood out for attempting to more clearly ground their design work in SDT \citep[e.g.,][]{andres2018super,VandenAbeele2015game}. 
Of particular note is work by Miller et al.~\citep{miller2019expertise}, who describe a process of `theory adaptation' to translate concepts from SDT (and Cognitive Load Theory) into generalisable game design features, with a view to redesign the citizen science game \emph{Foldit}. 
Competence, for instance, was first `operationalised' as ``the player's understanding of the tools available to them'' \citep[p.~5]{miller2019expertise}, and then instantiated as ``
giving the players more instruction on \emph{how} to use the tools available to them and less instruction on \emph{what} actions they should take
'' \citep[p.~6; emphasis added]{miller2019expertise}. 

Recognising that theoretical propositions may not readily translate into design \citep{gaver2012what}
, we had expected papers in this category to demonstrate more flexible applications of SDT -- more frequently, however, we observed more standard misinterpretations of the theory. 
Barata et al., for example, defined autonomy in gamified education as ``a sense of control over the learning environment'' \citep[p.~556]{barata2017studying}, a fairly common misconception that has prompted a number of clarifications within SDT\footnote{According to Deci and Ryan \citep[pp.~113-114]{deci1985general}, control refers to people's belief that \emph{outcomes} are controllable. Autonomy instead denotes one's sense of volition and endorsement of 
a behaviour. In other words, people may willingly engage in behaviours over which they have 
limited control \citep[e.g., mountaineering,][]{crockett2022self}, or feel pressured (i.e., low autonomy) to pursue activities whose outcomes they perceive considerable control over.} \citep[e.g.,][]{deci1985general}. A similar definition was provided by Tsai et al.~\citep[i.e. ``autonomy refers to a person’s sense of control over his or her own choices'';][p.~2]{tsai2021running}, whose work on a gamified social fitness system also misinterprets internalisation (cf. \autoref{OIT_diagram}) as referring to all forms of (a)motivation ``as the internalisation of motivation increases, one’s motivation can change from amotivation to intrinsic motivation'' \citep[p.~2]{tsai2021running}.

As in the analytic papers, when user studies failed to produce results in line with SDT, the theory itself was not questioned; rather, external factors or the artefact itself were identified as potential causes. After observing somewhat mixed results from their aforementioned design changes, for example, Miller et al.~lamented that their ``work was built on the existing structure of Foldit (now 11 years old)
'' \citep[p.~9]{miller2019expertise}. Notably, however, they also acknowledged that ``the [design] operationalisation process itself is a limitation in that there is a layer of abstraction between the theory and the implementation'', speaking to the issue of theory translation -- a concern also shared by industry practitioners. 

\section{Game Developers on SDT}
\label{gdc}
Outside the remit of academic games research, SDT has been referenced in several popular game design and gamification textbooks \citep[e.g.,][]{chou2019actionable,hodent2017gamer,schell2020art,schreiber2021game}, suggesting that SDT-based work presents a promising avenue for HCI games research to contribute to the games industry \citep{nacke2014games,nacke2016sigchi,nacke2018games,stahlke2020games}. Moreover, the theory could potentially figure as an effective `boundary object' \citep{star_this_2010} facilitating collaboration between academics and design practitioners \citep{velt2020translations,whitson2017voodoo}, despite differing aims and values \citep[e.g.,][]{gray2014reprio}. 
To form a more extensive view of the ways that SDT is understood and applied to games, we therefore examine how the theory has been discussed at a practitioner-facing venue -- the Game Developers Conference (GDC). 
GDC is the largest industry conference, attracting upwards of 15'000 attendees each year, and features talks and panels 
across a wide range of topics relevant to game design and production. Pertinent to 
our interest in theory translation, 
GDC and its affiliated events (e.g., GDC Europe) are widely considered a premier source of knowledge among games industry practitioners \citep{engstrom2019gdc,francis2021new}. Though commercial in nature, GDC resembles an academic conference in that speakers submit proposals that are reviewed by industry peers. Presentation 
materials from past events are typically made available on the GDC Vault, a subscription-based online repository. 


\subsection{Search and Analysis Procedure}
We searched the GDC Vault to identify content referring to SDT. However, as searching the many hours of video presentation records for relevant material was highly impractical, we narrowed our scope by first examining all available slide decks in the Design, Game Narrative, Independent Games, 
and Serious Games categories (N=897). We ran a keyword search of these documents on 7 May, 2020\footnote{Accessing the GDC Vault requires paid membership. Regrettably, we did not have access to the Vault for subsequent events.} for "motivat", "competen", "autonom", and "relatedness", excluding 742 presentations that contained none of these terms. The remaining 155 slide decks were downloaded and manually examined, resulting in the removal of a further 139 presentations for which there was no indication that concepts (e.g., motivation) were drawn from SDT \citep[e.g.,][]{jaffe2019cursed,wilson2011intentionally}. 
Across the remaining 16 presentations, our corpus encompassed 1316 slides and 14 hours 
of audiovisual material\footnote{Note that for Rigby's 2008 presentation \citep{rigby2008sustaining} only the slides and an audio recording were available.}. 

Next, the first author reviewed the material -- the textual slide content as well as the audiovisual record 
available on the GDC Vault -- 
and transcribed all SDT-related statements (N = 174, see supplementary material for transcripts and time stamps). The transcribed material was then analysed following an inductive process, where we first generated an initial set of codes to categorise the individual statements. Specifically, statements could pertain to (1) SDT in general, (2) SDT concepts, (3) examples of SDT concepts in games, (4) design considerations, and (5) miscellanea. In a final step, we collated
the coded statements into four broad themes describing how SDT has been used by game developers: as conceptual frame; as analytic lens; in folk theory; and critique of SDT.

\subsection{Findings}
\label{gdc_findings}

SDT was first introduced to GDC by Rigby in 2008 \citep[][along with a series of later talks, for which no slides were available in the GDC Vault, e.g., \citep{rigby2017freedom}]{rigby2008sustaining}. Speaking both as an academic and a consultant for Immersyve, Rigby emphasised the PENS model's 
scientific foundations as well as its utility for industry: ``if you put it (the PENS) in front of egghead academics, it'll stand up as science, but at the same time, it's practical
''. Indeed, SDT appears to have accrued considerable rhetorical power among games industry practitioners, as evidenced by some presenters referring to it as ``a popular theory in games at the moment'' \citep{lewis-evans2017throwing}, or otherwise commenting on the theory's prominence  \citep[e.g., ``we're going to use self-determination theory, because this is GDC, and I figure a lot of you are familiar with it'';][]{cox2018good}. 

Before we examine the ways SDT has been discussed at GDC, it is also worth noting that none of the reviewed presentations referenced SDT-based HCI games literature\footnote{Outside our corpus, Rigby's GDC 2017 talk \citep{rigby2017freedom} cites HCI games scholarship \citep[i.e.,][]{johnson2010personality,mcewan2014natural,wiemeyer2016player}, among other references, as evidence for the PENS model's validity.}, despite a shared interest in the theory and some industry collaborations among our academic corpus \citep[e.g.,][]{azadvar2018upeq,lehtonen2019movement}. This was somewhat surprising given HCI games reseachers' stated interested in contributing to the games industry \citep{nacke2014games,nacke2016sigchi,nacke2018games,stahlke2020games} and that HCI scholars have previously presented at GDC \citep[e.g.,][]{isbister2010better}. In fact, HCI games research was not entirely absent from our GDC corpus -- a presentation of UX practices at Epic Games \citep{hodent2014developing} referenced the GameFlow model \citep{sweetser2005gameflow}.

\subsubsection{SDT as Conceptual Frame}
Most GDC talks 
employed SDT as a \emph{conceptual frame to reason about what makes games engaging and commercially successful}. Rigby \citep{rigby2008sustaining} explicitly framed his presentation in these terms, emphasising that ``the underlying hypothesis [of the PENS model] is that you’re really going to have sustained fun and commercial success when you’re satisfying intrinsic motivational needs in your player''. Several presenters similarly focused on 
need satisfaction 
with a view to improve profit; for example, ``if your microtransaction just satisfies one of these needs, then you're doing great, and it's going to have value, and people are going to buy it'' \citep{cox2018good}. 

Among SDT concepts, 
autonomy was most 
prominent, 
with a total of 54 mentions \citep[including 8 mentions of ``agency'' in][]{hudson2011player}. Notably, Rigby referenced (to our knowledge) unpublished study findings 
that ``autonomy actually has the strongest relationship of all the needs to sustained subscriptions in MMOs'' \citep{rigby2008sustaining}. He further clarified that autonomy ``isn't exactly freedom [...] you can actually constrain people's choices as long as whatever path you've put in front of them, they feel that they've chosen to take'', emphasising its basis in an internal locus of causality. Other presenters, however, primarily framed autonomy in terms of ``having options'' \citep{burnell2012breaking}, ``meaningful choice and self-expression'' \citep{hodent2016gamer}, and making ``decisions that matter'' \citep{hudson2011player} -- terms that mirror those used in videogame marketing \citep{oliva2018choose}. 

Besides need satisfaction and the PENS model, intrinsic motivation was also considered important. Shirinian \citep{shirinian2014video}, for instance, emphasised that ``
if your game intrinsically motivates your players, they will want to play more often, they will be more deeply engaged [...] they will evangelise your product more, and in fact they're also learning themselves''. 
Others considered both intrinsic and extrinsic motivation core to the player experience. Hodent \citep{hodent2016gamer}, for example, indicated that ``intrinsic motivation is even more powerful [than extrinsic motivation] for long-term engagement'', but ``players' engagement is typically driven by extrinsic motivation, because they want to achieve external rewards, and avoid the opposite, typically the absence of rewards''. %
Similarly, some \citep{hodent2017gamer3, lewis-evans2017throwing} speakers discussed `non-contingent' (or in strict SDT terms, `engagement-contingent' \citep{deci1999meta}) rewards. Lewis-Evans \citep{lewis-evans2017throwing} described non-contingent rewards as ``participation awards'' that ``are less likely to be seen as controlling [...] because you don't have to do anything to get them''. Drawing from design examples from racing games (rubber-banding), MOBAs (passive resource generation), and 
`casual' games (catchup mechanics), Lewis-Evans further argued that ``non-contingent rewards can be quite useful for [...] balance, and making sure people experience more of your game''. 


\subsubsection{SDT as Analytic Lens} 
Rigby \citep{rigby2008sustaining} described the PENS model as ``a divining rod'' 
to assess ``how is this [feature] going to satisfy the player?''. Indeed, we observed a number of instances where speakers drew on SDT concepts to \emph{analyse particular game designs}. 
Hudson \citep{hudson2011player}, for instance, explained the effectiveness of \emph{Red Dead Redemption}'s \citep{rockstar2010red} climactic shootout scene 
through the player's learned competence. In particular, the analysis focused on the game's Deadeye mechanic, which provides brief invulnerability and slows time while the player selects targets. Hudson notes that ``as you play, you've probably gone into Deadeye hundreds of times ... 
if you were really good, you could maybe peel off five guys before they got to you''. Hence, in the final scene, when ``
the game forced you into Deadeye, and you're surrounded by twenty people, you panicked [...] you said `oh my god, my competence as a player, this isn't enough' [...] They sold that sense of despair and finality, through using that mechanic, better than any cutscene could have''.

Burnell instead used autonomy frustration to make a point about designing choices, contrasting ``a scene [in \emph{Portal}] \citep{valve2007portal} where you have to destroy an inanimate box'' with a similarly conflicting scene in \emph{Portal 2} \citep{valve2011portal2}. 
Noting that ``people love \emph{Portal}'s scene, and hate \emph{Portal 2}'s scene'', Burnell observed that ``in both [...] you have your autonomy taken away, but in [\emph{Portal}], the villain is making you do this heinous act, and in [\emph{Portal 2}], the game designer's making you do it [...] 
in one case, the player is angry with the villain, and in the other case, the player is angry with the game designer''.
 \vspace{-0.2em}

\subsubsection{SDT in Folk Theory}
We identified some cases where SDT concepts were used to develop a \emph{folk theory} \citep{gelman2011concepts}; in other words, \emph{where SDT concepts 
were adapted into developers' own conceptions 
of player experience and good game design}. As with other instances where psychological theory has been applied to game design practice \citep[e.g., the big five personality model in][]{vandenberghe2016engines}, these folk theories 
vary in terms of how faithfully they reflect 
SDT tenets. Griesemer \citep{griesemer2011design}, for example, described \emph{Halo}'s \citep{bungie2001halo} dodging and cover mechanics in terms of competence and autonomy, respectively, to emphasise the contextual value of need satisfaction, claiming that ``competence in a shooter, specifically dodging, is much more fun when you're under low pressure, and autonomy is much more fun when you're trying to tackle a problem that's much more difficult''. In this way, he argued that ``when games satisfy these complex needs, that's when they're really fun. And this complexity comes from the context that the needs are being satisfied in''. 

Lewis-Evans \citep{lewis-evans2017throwing} instead discussed common industry practices for supporting intrinsic motivation, noting that ``a lot of companies [...] talk about intrinsic motivation [in terms of] goal structures, and adding a bunch of stuff on top of the game''. Acknowledging the merits of such approaches, Lewis-Evans championed good game feel \cite{swink2008game} as ``the ultimate'' factor. 
In this view, ``games that are commonly cited [as intrinsically motivating] like \emph{Destiny} \citep{bungie2014destiny} [...] 
just feel satisfying to play. The action of just moving a controller or a mouse feels impactful and immersive''. 
\vspace{-0.2em}

\subsubsection{Critique of SDT}
Finally, we noted a few instances where game developers \emph{critiqued 
the extent to which need satisfaction 
could meaningfully speak to game design}. Though these critiques were somewhat facile in their engagement with theoretical tenets, they form notable counterpoints to perspectives that leverage SDT as a conceptual frame. Burnell \citep{burnell2012breaking}, for instance, questioned whether \emph{Farmville} \citep{zynga2009farmville} could really be considered ``the pinnacle of game design'', even though ``you can plant crops in any pattern you want, giving you autonomy. The only skill is clicking, so no-one feels a lack of competence. And your friends are relying on you to play, and benefit from you playing, giving you relatedness''. 

In another example, Griesemer \citep{griesemer2011design} was critical of the relevance of SDT-based study findings for game design practice, with respect to an experiment \citep[Study 4]{przybylski2009motivating} where the researchers ``modded \emph{Half-Life} \citep{valve1998half} [...] one version [...] was [violent] and [the other] was non-violent; it was sort of like a game of tag''. Although some details of the study were misconstrued\footnote{But note that these errors appear to originate with the relevant secondary source \citep{madigan2010psychology}. In particular, the study is claimed to have found that both versions of the game satisfied basic needs (which was not tested). Instead, the study finds that violent content did not significantly predict enjoyment or future play preference, while competence and autonomy accounted for a similar degree of variance 
irrespective of the version of the game \citep{przybylski2009motivating}.}, his conclusion likely echoes industry sentiments: ``we all know that \emph{Tag of Duty} would not sell''.

\section{Discussion}
\label{Discussion}

Self-determination theory (SDT) figures prominently in Human-Computer Interaction \citep{ballou2022self}, and HCI games research in particular \citep{tyack2020self}. 
Examining its applications helps shed light on the ways theory is -- and could be -- used in HCI. In light of SDT's wide-reaching influence, it may also clarify why some theories get taken up in research and design practice, while others fail to gain traction. 
In this paper, we have reviewed a wider range of HCI games literature informed by SDT, examined games research conducted by SDT scholars, and identified industry perspectives on the theory. 

First, we found that descriptive theory uses account for the bulk of 
citations in the HCI games corpus (n=142, 54.83\%, \autoref{tab:codes}). These works appear largely uninterested in engaging SDT beyond its rhetorical power \citep{halverson2002activity} or employing SDT-based measures. In contrast, SDT-informed hypothesis testing -- the most fruitful application of SDT 
according to Rigby and Ryan \citep{rigby2011glued} -- featured in a comparatively small subset of papers (n=44, 16.99\%). 
Second, a fraction of SDT concepts -- namely, need satisfaction and intrinsic motivation -- dominate the literature (\autoref{theory_concept}), with other concepts and mini-theories largely overlooked. 
Third, misconceptions and unusual interpretations of the theory can be found 
across all levels of theoretical engagement. 
Finally, across the corpus, we identified a reluctance 
to critique or contest claims from SDT (
\autoref{tab:codes}), even when study results seemed to represent evidence against the theory. Instead, divergent results were consistently explained by recourse to other theories, or dismissed as potential quirks of study design. While stronger theoretical engagement 
was more commonly observed in our expanded corpus -- for example, in works where SDT informs hypothesis derivation and the interpretation of results \citep[e.g.,][]{born2021motivating,burgers2015feedback,mills2020self} -- few such works `talk back' to SDT \citep{beck2016examining}, 
either to extend or challenge the theory. Collectively, these findings underscore our prior observation \citep{tyack2020self} that HCI games research presently figures SDT less as a \emph{theory}, and more as a \emph{paradigm} \citep{rogers2012hci}, i.e., ``a set of practices that a community has agreed upon, (including) questions to be asked and how they should be framed; phenomena to be observed, how findings from studies are to be analyzed and interpreted'' \citep[p.~4]{rogers2012hci}. 

What about the translational value of SDT-based games research for design practice? The 
work we reviewed can largely be considered `applied research' \citep{colusso2019translational} -- situated between basic research (as in core SDT scholarship) and game design practice. In this sense, HCI games research could be well-positioned to bridge a key part of the theory-practice gap \citep{velt2020translations} and provide resources for use in design practice \citep{colusso2017}. At GDC, however, 
HCI research involving SDT was entirely absent. 
This is at odds 
with HCI games scholars' stated interest in industry engagement \citep[e.g.,][]{stahlke2020games}, and 
assertions that HCI research has practical relevance to game development (e.g., implications for design). From our analysis of the GDC presentations, however, it is clear that game design practitioners \emph{do} value 
SDT research for their work. It is also clear that SDT scholars themselves \citep[e.g.,][]{rigby2008sustaining,rigby2011glued} have worked to cultivate this interest in the theory among game developers. 
At the same time, we identified a willingness among games industry practitioners to adapt or 
disregard theoretical tenets 
\citep[e.g.,][]{burnell2012breaking,griesemer2011design}. This stands against the somewhat indifferent and deferent attitude we observed in the majority of the academic corpus
, where results that are potentially inconsistent with SDT were explained in ways that foreclose re-evaluation of the theory. 
In this way, HCI games research eschews critique of SDT, and neglects to question 
the extent to which its tenets represent immanent truth. 



This leaves us with several questions: If SDT holds much promise for advancing HCI games research, as we have previously argued \citep{tyack2020self,tyack2020masterclass}, why do HCI games scholars engage so little with its theoretical tenets? 
And if we do not effectively leverage SDT for research or to inform industry practice, how did SDT nevertheless attain considerable popularity in the community? 
In the following, we comment on issues with existing SDT games scholarship, examine conditions underlying the theory's current use as paradigm, 
and identify avenues for more fruitful SDT-based HCI games research. 

\subsection{Issues with SDT Games Scholarship}
\label{issues}

SDT scholarship is much cited in games research, and in many instances forms a core part of a study's foundational literature. This has been widely beneficial for HCI games research \citep{tyack2020self} -- it has provided 
methods for assessing the player experience (i.e., PENS and IMI), as well as a shared vocabulary for discussing games and play across a variety of topics. 
However, future SDT-based games research would also benefit from a more measured perspective towards these works, particularly regarding concepts and claims that are absent from, or inconsistent with, the wider theory. 

Examining SDT scholarship on games (\autoref{games_sdt}) may also help explain trends in the ways that HCI games research has deployed the theory (\autoref{Results}). 
Notably, several SDT games 
publications \citep[e.g.,][]{przybylski2009motivating, przybylski2014competence,ryan2006motivational} report on multiple studies -- reflecting journal standards in psychology -- which arguably left relatively little room for 
elaborate discussions of the theory. The often perfunctory treatment of SDT in HCI games research, then, may at least partially reflect its predecessors' salient features; namely, an emphasis on basic need satisfaction, enjoyment, and their demonstrably effective self-report measures (i.e., PENS and IMI, respectively). 

Further, our analysis showed that SDT mini-theories are seldom mentioned in HCI games research (see section \ref{prevalence}). Similarly, SDT games scholarship has rarely leveraged 
these theories to inform conceptual or empirical work. The role of motivational regulation for play, for example, has received little attention, despite 
organismic integration theory (OIT) providing testable propositions \citep[e.g.,][]{ryan2017organismic} 
and the availability of a validated OIT-based instrument to assess gaming motivation \citep{lafreniere2012development}. Instead, SDT tenets have been sequestered into a game-specific category (as described in \autoref{games_sdt}) that incorporates the need density hypothesis, 
the hero construct, and the PENS model \citep{ryan2017motivation}. However, many of these concepts' theoretical links remain tenuous and often lack empirical support. 

In the case of the need density hypothesis, SDT scholarship is yet to adequately clarify why videogame play 
satisfies basic needs more effectively than other activities 
\citep{ryan2017motivation}. %
Moreover, it is not clear that the need density hypothesis explains any more about the relationship between game overuse 
and wellbeing than existing SDT models \citep[e.g.,][]{milyavskaya2013psychological, vallerand1997toward}, which 
%
emphasise that ``[cognitive, affective, and behavioural] consequences are of the same level of generality as the level of generality of the motivation that engendered them'' \citep[p.~276]{vallerand1997toward}. Importantly, empirical work supports the notion that global wellbeing is \emph{primarily} affected by need satisfaction 
\emph{in daily life} \citep[e.g.,][]{allen2018satisfaction,johnson2022unsatisfied, milyavskaya2013psychological}\footnote{Although Allen and Anderson \citep[]{allen2018satisfaction} interpret their study in terms of the need density hypothesis, their findings only represent support for the claim that global need satisfaction predicts global wellbeing more effectively than in-game need satisfaction.
}. 
In suggesting that games themselves contribute to overuse, the need density hypothesis stands apart from other SDT theorisations of maladaptive behaviour, which instead point to ``the thwarting of autonomy, through either excessive control, conditional regard, or lack of empathy'' as a causal factor that ``leads to dysregulation and ill-being'' \citep[p.~386]{ryan2016autonomy}. Put differently: it is difficult to square 
Rigby and Ryan's analysis, which draws equivalences between videogames and sugar in causing over-indulgence \citep[p.~104-105]{rigby2011glued}, with SDT tenets that primarily locate the basis of eating disorders in the wider ``role of familial factors in setting up the dynamics of introjection and internal control'' \citep[p.~419]{ryan2016autonomy}. 

Similarly, Rigby's articulation of the hero construct \citep{rigby2009virtual,rigby2017time} 
incorporates contextual autonomy support \citep{deci1987support} and intrinsic goal pursuit \citep{ryan2017goal}, though 
it remains unclear to what extent it presents a meaningful extension of these concepts to games. To our knowledge, no empirical studies have investigated whether a heroic framing explains games' motivating qualities better than existing tenets on contextual need support \citep{gillet2010influence,haerens2015perceived}. It is also worth noting that heroic narratives have seen substantive critique in game design \citep{dena2017finding,jayanth2016forget,nicklin2022writing}, particularly in that they de-emphasise relatedness \citep{reay2022videogames} by casting other players and characters as resources to be mined for rewards and acknowledgment \citep{jayanth2016forget,nicklin2022writing} -- notions more reminiscent of external regulation \citep{ryan2017organismic} and external goal pursuit \citep{ryan2017goal}. 
Further, the hero construct is unlikely to productively inform the design of many popular 
game genres (e.g., horror) or formats (e.g., multiplayer games), as ``heroism is not a universal desire for play'' \citep[p.~34]{dena2017finding}. To suggest that heroism can underpin a \emph{general} framework for understanding virtual worlds \citep{rigby2009virtual} or entertainment media \citep{rigby2017time} seems 
an overreach. 

That said, the need density hypothesis was almost entirely absent from our HCI games corpus and the hero construct was mentioned only in Rigby's GDC presentation \citep{rigby2008sustaining} -- suggesting that these aspects of SDT games scholarship have had limited 
influence on research and design practice (but see section \ref{needswell} for more recent work). 

The PENS model (section \ref{pens}), in contrast, featured prominently both in the academic and the industry corpi. Some scepticism is nevertheless in order. Although the PENS model draws from basic psychological needs theory \citep{rigby2011glued,ryan2006motivational}, SDT games scholarship presently lacks concrete propositions that would explicate the interplay between PENS constructs, as well as their relation to game design, motivation, or wellbeing. This 
is particularly evident with regards to intuitive controls and immersion, whose theorised relation to need satisfaction inexplicably vary from study to study \citep[i.e.,][]{przybylski2012ideal,przybylski2014competence,ryan2006motivational}, and whose links to the wider theory remain unexplained. 
Indeed, 
relatively few papers in our HCI games corpus (n=37; 14.29\%) 
measured immersion using the PENS. Even fewer \citep[i.e.,][]{kim2015sense,smeddinck2016difficulty,weech2020narrative} formed hypotheses involving the measure -- and those that did relied on literature \emph{outside} SDT as the basis of their predictions (e.g., flow theory). 
Counter to assertions in SDT scholarship to ``drive real hypothesis testing'' \citep[p.~167]{rigby2011glued} and ``epistemological coherence and rigour'' \citep[p.~113]{ryan2019brick}, the PENS model 
presently resembles -- to quote its creators -- 
a ``divining rod'' \citep{rigby2008sustaining} that risks ``inviting pure speculation about causal connections that might not actually exist'' \citep[p.~168]{rigby2011glued}. 
In this way, SDT games scholarship may have 
served as a blueprint for 
the relative paucity of theory-informed research (e.g., SDT-based hypotheses) we identified in the HCI games corpus, as well as 
legitimised cursory and tenuous applications of the theory. 
Similarly -- and to some extent perhaps as a consequence -- the empirical basis for many claims in SDT games scholarship is dubious. 
For example, it has been claimed that the PENS scale was validated \citep[e.g., as stated in][p.~246]{przybylski2009motivating}; however, this analysis has never been published\footnote{Notably, the Immersyve website names an independent study from HCI scholars \citep[]{johnson2010personality} as the `PENS Validation Study' \citep[]{immersyve2011pens}, indicating that SDT scholars are \emph{aware} of HCI games scholarship -- yet these contributions are unacknowledged in published research and other official channels \citep[e.g.,][]{center2020virtual}.}. 
In fact, independent validation studies from our HCI games corpus \citep{denisova2016convergence,johnson2018validation} suggest that intuitive controls and competence might not actually represent distinct constructs. Likewise, SDT games scholarship features several unfounded claims about the relation between game design and need satisfaction. Rigby and Ryan, for example, link optimal challenge to higher competence satisfaction \citep{rigby2011glued}; a claim for which recent HCI games research found no empirical support for \citep{deterding2023objective}. Similarly, that autonomy can be effectively satisfied by games that allow players to enact their avatar's identity \citep{rigby2011glued}, or that offer a high degree of choice \citep{rigby2011glued}, has never been empirically tested in SDT games scholarship. Indeed, several of the reviewed HCI games studies have examined and complicated these claims, indicating that identification with one's avatar mediates the impact of avatar customisation on autonomy need satisfaction \citep{Birk2016fostering}, or that, for some players, a high degree of in-game choice may actually be perceived as controlling \citep{tyack2021small}. 

Yet few HCI games papers `talk back' to SDT -- to compare their findings, challenge existing claims, or to extend theoretical tenets. Likewise, although SDT positions itself as a broad theory that operates in similar ways across application domains \citep{ryan2017introduction,ryan2019brick}, SDT games scholarship has to our knowledge rarely, if ever, informed the wider theory. 
Immersion and intuitive controls, for instance, have featured only in SDT research into games, despite claims that ``[the PENS] framework is agnostic to any specific technology or design'' \citep[p.~172]{ryan2020motivational}, indicating that these constructs could also be productively applied to SDT-based studies of other entertainment media \citep[e.g., TV viewing experiences;][]{adachi2017wait} or technology use \citep
{peters2018designing}. 

Ultimately, SDT is underutilised in HCI games research, but also rarely questioned. 
While both issues are to some extent 
due to limited engagement with SDT scholarship 
\citep[and tendencies within HCI to gloss over prior work,][]{marshall2017misrepresentation, marshall2017throwaway,marshall2017little}, they likely also stem from the looseness of SDT games scholarship, which 
likewise contains limited use, development, and critique of SDT at large. As such, continued use of SDT as an \emph{unquestioned paradigm} -- i.e., an unquestioned ``set of practices 
[... including] questions to be asked and how they should be framed; phenomena to be observed, how findings from studies are to be analyzed and interpreted'' \citep[p.~4]{rogers2012hci} -- risks impeding HCI researchers' capacity to develop our knowledge of player experience, motivation, and wellbeing. It also poses a missed opportunity to engage more deeply with game design practice, despite the shared interest in SDT. 


\subsection{From Unquestioned Paradigm to Intentional Theory Use}

Thus far we have established that HCI games research presently figures SDT less as a theory, and more as a paradigm \citep{rogers2012hci}. 
%
We also outlined how SDT games scholarship may have in part contributed to the theory being deployed in this way. 
However, this does not account for \emph{why} SDT has been readily taken up as a paradigm. 
In the following, 
we reflect on the theory's 
reception in HCI games research and among industry practitioners, and how this relates to the ways SDT has been put to use. 
We then discuss how HCI games researchers may move beyond SDT as unquestioned paradigm, and instead
leverage the theory towards more fruitful ends. 




\subsubsection{Why Self-Determination Theory?}
\label{why}


The HCI literature has proposed several attributes of `good' theory, for example, precise counterfactual propositions \citep{oulasvirta2022counterfactual} or generative capacity for design \citep{beaudouin-lafon2021generative,hook2012strong}. Prior work has also speculated about the factors that inhibit theory use in HCI, such as iterative design practices \citep{oulasvirta2022counterfactual}. 
Rarely, however, has the literature considered what makes a theory `successful', that is, what conditions facilitate its uptake in a community of practice \citep[e.g., HCI researchers, industry professionals]{rogers2004new,rogers2012hci} -- conditions presumably also advantageous to establishing (a specific) theory as a motor theme \citep{kostakos2015big}. 
SDT stands out for 
its widespread adoption 
across research areas and application domains, including a growing presence in HCI \citep{ballou2022self}. Yet for all assertions that ``SDT stands unrivalled in its popularity among HCI researchers
'' \citep[p.~261]{poeller2022self}, the reasons for this remain at best implicit. 
%
Examining SDT's popularity in games may therefore help elucidate the circumstances under which theory comes to shape research and design practice 
(e.g., as paradigm), 
and why sometimes even good theories get relegated to the sidelines. 
It 
may also help explain why SDT has been privileged over alternate theoretical perspectives in games research \citep{poeller2022self,tyack2021offpeak}. The following observations should not be understood as causal claims per se
, but rather serve 
to spotlight taken-for-granted 
aspects of theory use. 

Critiquing the primacy of SDT in HCI games research over other motivational frameworks, 
Poeller and Phillips muse ``
whether some researchers may be defaulting to SDT because they feel it is the only option available'' \citep[p.~261]{poeller2022self}. We argue that this is no coincidence. 
SDT scholars have actively cultivated interest in the theory through multiple lines of translational work \citep{colusso2019translational}. This is perhaps most evident in the PENS model, which represents 
a translation of basic psychological needs theory and (to a lesser extent) cognitive evaluation theory to 
games. Compared to other motivational theories that still await such specification \citep[but see work on GameFlow,][]{sweetser2005gameflow,sweetser2020gameflow}, 
SDT concepts 
may appear more intuitive 
and relevant in the context of games. 

With need satisfaction and intrinsic motivation, in particular, SDT put forward 
a conceptualisation of `good' player experience (PX) that is 
applicable to many areas of games research (e.g., design, wellbeing, applied games), without necessarily requiring deep engagement with SDT tenets. With regards to HCI theory, Rogers similarly noted that 
``stand-alone, one-off terms that conjure up what they mean intuitively have been the most widely taken up -- even though they are often used much more loosely and in underspecified ways'' \citep[p.~83]{rogers2012hci}. Arguably, it is precisely this looseness that renders said SDT concepts more intuitive yet prone to misconceptions, more amenable to descriptive applications in research, and which facilitate their adaptation into industry folk theories \citep[e.g.,][]{griesemer2011design,lewis-evans2017throwing}. 





The PENS and IMI scales have likewise played a major role 
in popularising SDT 
in games research, providing a means to operationalise SDT's conceptualisation of good PX. Indeed, the majority of papers in our corpus deployed SDT-based measures (54.05\%, n=140, see also \autoref{tab:measures}), 
and 
frequently with no apparent theoretical rationale (n=76). Although this is emblematic 
of the often perfunctory treatment of SDT in the literature, it also suggests that these scales readily ``graft onto existing practice'' \citep[p.~84]{rogers2012hci} of HCI games scholars, who 
rely on self-report instruments for evaluation and to study the player experience \citep{johnson2018validation,vandenabeele2020development}. 
In this way, 
the PENS and IMI questionnaires \citep[and the long-standing absence of effective alternatives;][]{law2018GEQ} 
rendered SDT more 
accessible to HCI games research, but also helped cultivate the conditions for shallow engagement with the theory. It remains to be seen whether descriptive-methodological citations, where SDT does not otherwise inform the citing work, will decline 
now that alternate (i.e., non-SDT-based) validated PX questionnaires are openly available \citep[e.g.,][]{vandenabeele2020development}. 

%
It is worth noting that initial SDT games scholarship \citep{ryan2006motivational} 
not only 
introduced SDT concepts and measures 
to games 
research, but also 
functioned to displace competing approaches \citep[e.g.,][]{anderson2004update,bartle1996hearts,yee2006demographics}. For instance, showcasing that the PENS model 
%
exhibits improved incremental validity 
\citep{ryan2006motivational} 
over Yee's gaming motive framework\footnote{The PENS and Yee's framework are arguably incommensurable, as the former is conceptualised as a \emph{universal} model of player motivation \citep{ryan2006motivational}, while the latter 
represents a \emph{differential} account \citep{yee2006motivations}. A more apt comparison would be to test the PENS against other universal need theories \citep[e.g., akin to][]{sheldon2001satisfying}.} \citep{yee2006demographics,yee2006motivations}, as well as implying that the latter ``largely reflect(s) the structure and content of current games'' \citep[p.~348]{ryan2006motivational}. These critiques were reiterated and indirectly amplified in subsequent publications \citep[i.e.,][]{rigby2011glued,przybylski2010motivational}, such as in 
calls for `clear' and `good' theory to drive `real' hypothesis testing \citep[p.~167-168]{rigby2011glued}. Of course, such statements may chiefly reflect common academic writing practice \citep{giner2012science} rather than 
intent to disparage other frameworks. Indeed, they represent a (somewhat implicit) example of `talking back' to theory \citep[][see next section]{beck2016examining}. Nevertheless, they likely helped further 
cement SDT's status as a `default' theory \citep[p.~261]{poeller2022self} in games research. 

More speculatively, some of these critiques are perhaps also 
a product of 
competing commercial interests\footnote{In 2015, Yee co-founded his own market research company, Quantic Foundry, which focuses on player motivation.} to better 
position Rigby and Ryan's UX consultancy, Immersyve. 
%
Indeed, SDT scholars actively fostered industry engagement via several translational efforts \citep{colusso2019translational}, such 
as GDC talks \citep[e.g.,][]{rigby2008sustaining,rigby2017freedom}, a popular science book \citep{rigby2011glued}, and materials \citep{rigby2007player} made available on the Immersyve website. Importantly, our analysis of GDC talks suggests that these efforts did not simply serve dissemination purposes, but in introducing SDT as a conceptual frame to reason about sustained engagement and profit \citep[e.g.,][]{rigby2008sustaining} directly catered to practitioner concerns \citep{kanter2005theories}. 
Industry practitioners, in turn, appear to reference 
SDT 
for the presumed 
rhetorical power \citep{halverson2002activity} it conveys, with presenters perhaps citing the theory to legitimise existing intuitions \citep[e.g.,][]{griesemer2011design,hudson2011player} 
-- indeed, the common assumption at GDC 
is that game designers have already heard of SDT \citep{cox2018good,lewis-evans2017throwing}.
Moreover, as a professed scientific theory \citep{rigby2008sustaining,ryan2017self,ryan2019brick} backed by five decades of 
research, 
SDT research 
may appear more legitimate to game developers 
than comparably nascent HCI games scholarship.

Finally, ``the reception of a theory is shaped by the extent to which a theory resonates with the cultural presuppositions of the time
'' \citep[p.~394]{dimaggio1995comments}. 
This also applies to SDT games scholarship, where much of its conceptual apparatus -- immersion, intuitive controls, heroic frame; but also PENS notions of competence and autonomy that centre mastery and choice -- 
echoes mainstream game design conventions 
\citep{kultima2009casual,oliva2018choose}. 
It is not clear whether this simply reflects SDT scholars' personal views on games
, or whether this is the product of concerted efforts to 
establish industry relevance. 
The latter might account for 
why the hero construct has been incorporated into SDT games scholarship, not based on empirical evidence, 
but 
because it is ``[w]hat everyone else 
refers to
'' \citep[p.~35]{dena2017finding}; whereas other SDT concepts (e.g., %
functional significance, 
continuum of extrinsic motivation) -- which in principle could fruitfully expand, 
nuance, and \emph{complicate} mainstream conceptions of game engagement -- have yet to be considered. 

But note that theories convey normative meanings \citep{gergen1978toward,ghoshal2005bad,zoshak2021beyond}, with potentially far-reaching consequences when applied to technology design: 
``As its implications and applications are borne out, every theory becomes an ethical or ideological advocate'' \citep[p.~1354]{gergen1978toward}. 
By championing immersion and intuitive controls 
as hallmarks of need-satisfying play, for instance, 
SDT scholars imbue them with 
scientific authority. In this way, SDT games scholarship does not simply reinforce dominant design values, but risks marginalising 
perspectives on worthwhile game design that run counter to 
SDT-informed notions of good PX \citep[e.g.,][]{schmalzer2020janky,tyack2021offpeak}. A more overtly disconcerting example can be found in Rigby's recommendation to GDC attendees \citep{rigby2008sustaining} that ``you can actually constrain people's choices as long as [...] 
they feel that they've chosen''. 
This may not be detrimental to players' experience; indeed, 
perceived autonomy 
was linked to sustained engagement and future play intention \citep{rigby2008sustaining,ryan2006motivational}. However, it raises questions about the commodification of need satisfaction \citep[see also][
]{soderman2021against} and the ethical ramifications of autonomy manipulation \citep{bennett2023does,nguyen2020challenges} -- especially when linked to a theory nominally concerned with 
self-determination. 

HCI games research is likewise 
embroiled in (perpetuating) these cultural and ideological presuppositions. 
In their study of games' need-restoring qualities, for example, 
Tyack and Wyeth 
\cite{tyack2021small} -- acknowledging their complicity -- 
question whether SDT's view of wellbeing, as a product of self-realisation
\citep{ryan2017economic}, over-privileges individual responsibility \citep[see][]{kou2019turn,whitson2013gaming} 
and ``enroll[s] play into practices of responsible self-governance
'' \cite[p.~19]{tyack2021small} -- and correspondingly de-emphasises the state's role in providing affordable public health care, the `safety net' that SDT otherwise endorses\footnote{Note that neoliberal tenets of individual governance and self-optimisation also feature in SDT's ideal welfare system, as exemplified by the following quote: ``To function as an effective support within capitalism, however, a safety net must be set at an optimal level---not so high as to discourage people from undertaking productive tasks that they might otherwise not be motivated to do [...] An optimal range would have the safety net set high enough to functionally support a life, but not so high that it crowds out meaningful incentives and personal initiative for entry-level labor'' \citep[p.~609-610]{ryan2017economic}.}. 

SDT has been useful for HCI games research \citep{tyack2020self}, it is indeed ``popular for a reason'' \citep[p.~261]{poeller2022self}. However, this popularity belies a legitimacy trap \citep{dourish2019user} that has allowed perfunctory applications of the theory to proliferate and lends unwarranted credibility to SDT games scholarship. HCI games researchers cite SDT because it is frequently mentioned in HCI games papers\footnote{
Our previous review \citep{tyack2020self} is often cited \emph{specifically} to support this point.}, because questionnaires are available, or because industry practitioners reference the theory; rarely 
to purposively leverage its theoretical tenets. 
Even scholarship critical of the primacy of SDT in games claims that the theory has been subject to ``widespread validation'' in HCI \citep[p.~261]{poeller2022self} -- a misconception that the present work has hopefully dispelled -- 
and in this way inadvertently 
perpetuates the very paradigm 
it intends to challenge.

\subsubsection{Moving beyond the Paradigm}
\label{moving}

Continued deployment of SDT \emph{as a paradigm} 
impedes HCI games research in several ways: 
It implicitly guides what phenomena are considered and how they are studied \citep{rogers2012hci}, eschewing more reflexive and transparent uses of SDT. It rarely involves `talking back' to theory \citep{beck2016examining}. 
It may also narrow the ways we think about games and play \cite{tyack2021offpeak}. 
Further, it may conceal 
theoretical inconsistencies and ideological underpinnings, and 
crowd out alternate theoretical perspectives -- reinforcing use of SDT as an \emph{unquestioned paradigm}. 
Lastly, in the absence of translational work in HCI games research, 
design practitioners are unlikely to notice or benefit from HCI games scholarship. 

As indicated by our analysis of GDC presentations, game designers are already familiar with many of SDT's core concepts. 
Yet theory is often underspecified for design \citep{gaver2012what,rogers2004new}, requiring further translation, and many claims in SDT games scholarship lack empirical support. 
The analytic and generative uses of theory we observed in our corpus suggest that, in principle, 
HCI games research is well placed 
to contribute to game development practice -- by 
testing, refuting, 
or extending theoretical propositions; and translating SDT concepts to design. 
However, 
success in this endeavour depends, to some extent\footnote{Forming stronger links between HCI games research and game development practice would also benefit from more engagement with flagship games industry events, such as GDC. Of course, there are practical reasons why GDC in particular is not widely attended by HCI scholars; for example, attendance is expensive, and costs are unlikely to be reimbursed by academic institutions; convention dates may coincide with teaching or other mandatory activities; and travelling to the U.S. may be considered unsafe or environmentally wasteful. The active participation of HCI games scholars -- for example, through presentations, panels, and so on -- represents a clear opportunity to promote their SDT-based research and its benefits. Although we remain somewhat pessimistic that funding systems will begin to support greater academic participation at game development events, researchers may include event attendance in grant applications, for example, as a means of industry outreach, particularly for projects where translation of theoretical HCI knowledge is either the object of study or one the expected outcomes \citep[e.g., see][]{velt2020translations}.}, 
on whether HCI games scholars go beyond the currently dominant applications of the theory -- namely, perfunctory discussions of need satisfaction and intrinsic motivation (\autoref{theory_concept}) -- and %
further 
engage with the implications of SDT's numerous theoretical tenets, and existing findings derived from relevant HCI games research. 

One way forward is to 
investigate unsubstantiated claims in SDT games scholarship %
to establish links between 
SDT concepts, game design, and 
behavioural and psychosocial outcomes. 
That said, most existing propositions 
\citep[e.g., regarding 
optimal challenge and competence, see][]{deterding2023objective} are insufficiently specified 
to 
support 
hypothesis-testing \citep{scheel2022most}. Hornbæk \citep[p.~188]{hornbaek2018commentary} has identified a parallel issue in usability research, which 
``contains very few propositions, that is, rarely commits to clear statements regarding the relations among dimensions
''. 
The near total absence of propositions about SDT concepts and games-related phenomena 
is especially jarring given that the wider SDT literature \citep[e.g.,][]{ryan2017self} already provides fairly well-articulated propositions. 
Delivering and investigating such propositions (see \autoref{avenues}) would 
facilitate progress in HCI games scholarship, in the form of a cumulative body of research, and allow study findings to `talk back' 
to SDT\footnote{We are less optimistic about the prospect of HCI games research actually influencing SDT scholarship. 
However, recent collaborations between HCI and SDT scholars \citep{ballou2024basic} represent a promising step in this direction.}. 

%
What does `talking back' to SDT entail? 
We apply here Beck and Stolterman's models of theory use \citep[][see also \autoref{hci}]{beck2016examining} to compare two examples from our corpus \citep[i.e.,][]{johnson2015base,tyack2021small}
. 
Both papers 
feature `self-determination theory' as author keywords, suggesting that the theory figures prominently in these works: 
%
Johnson et al.~\citep{johnson2015base} explored variations in PX across different game genres and observed that need satisfaction 
(as measured by the PENS) 
does not adequately account for the appeal of massive online battle arena (MOBA) games. 
Their findings are explained in terms of genre-specific properties and that the PENS measure does ``not fully capture the components of PX that attract people to play MOBA games%
'' \citep[p.~2270]{johnson2015base}. 
%
%
While the divergent findings present 
opportunity for 
`talking back' to theory, 
the authors leave the 
incongruity between SDT tenets and MOBA games largely uncommented\footnote{But note that as a CHI 2015 paper, Johnson et al.~\citep{johnson2015base} had to operate within a strict 10-page limit (incl. references) and that the Games and Play subcommittee did not yet exist, which may have curtailed opportunities for `talkback'.
}. 
In this way, the work leaves unclear how (if at all) SDT could inform future work, and (inadvertently) cements its use as paradigm. 

Tyack and Wyeth's investigation of autonomy-supportive play \citep{tyack2021small}, in contrast, %
%
leveraged SDT 
to derive study hypotheses 
and analyse interview data. The authors `talk back' by comparing their findings 
to SDT literature, for instance, noting that ``it is surprising that the substantial changes observed in autonomy satisfaction did not predict changes in happiness or calmness. The theory itself is somewhat unclear on this point [...] identifying the circumstances that produce these `typical' cases [...] 
seems essential to formulate adequate hypotheses'' \citep[p.~18]{tyack2021small}. In this way, the work raises open questions for future research and calls attention to opportunities for theory development. 

Finally, we found few design applications of SDT in our academic corpus (\autoref{tab:codes}), and some authors commented on difficulties \citep[e.g.,][]{miller2019expertise} with translating the theory. 
Work
on intermediary knowledge building in HCI \citep[e.g.,][]{beaudouin-lafon2021generative,ploderer2021diagramming,vries2023blueprints} may offer some pointers for linking SDT and design in a more principled manner, 
for instance, by 
formulating generative questions based in theoretical concepts \citep[][see also \autoref{avenues} for preliminary examples]{beaudouin-lafon2021generative}. 
These initiatives are worthwhile as a means to 
further develop theory \citep{colusso2019translational,stolterman2010concept}, build on existing HCI games research, and improve game development practice. 
Moreover, our analysis of GDC presentations indicates that industry practitioners' experiences of applying SDT in 
design 
can also create opportunities to (con)test theoretical tenets. Even if these 
claims are yet to be formally tested via academic studies 
\citep[e.g.,][on contextual need satisfaction]{griesemer2011design}, 
they represent serious engagement with theory and its limits, and 
form fruitful opportunities for further academic investigation \citep{colusso2019translational,gray2014reprio}. 

Another promising avenue for bridging %
theory and practice is to develop translational resources for design \citep{colusso2017,yoo2023beyond} -- informed by SDT and grounded in relevant HCI games research. 
In light of the ways that game designers themselves appear to conceptualise SDT in their own work, translating the theory's tenets into more useful resources for 
design practice may require some degree of flexibility. This is not to say that we should celebrate work that diverges extensively from SDT tenets, but rather that strict theoretical fidelity may not be the \emph{sine qua non} of translation \citep[see also][]{velt2020translations}. For example, although need frustration is unequivocally considered a negative experience in SDT \citep{ryan2017basic}, it may have other applications in PX design \citep[e.g., enhancing players' emotional responses;][]{burnell2012breaking,hudson2011player}. 
Further, translating 
HCI games research into alternative formats \citep[e.g., online articles, podcasts, GDC presentations;][]{colusso2019translational, kunzelman2021game,ramsden2015user} that are more amenable to design practitioners may require shedding some of the formal rigidity that has given rise to structural norms and standards in HCI (games) literature \citep[e.g., see][]{carter2014paradigms, dourish2006implications}.

\section{Avenues for SDT-based HCI Games Research}
\label{avenues}

HCI is among the primary domains in which SDT continues to inform games research. 
This paper has thus far identified gaps and challenges in SDT-based HCI games research, and argued for more intentional theory use. Although we 
briefly outlined some potential research avenues in our previous review \citep{tyack2020self}, we lacked the space in that paper to make specific calls for action. Here, we more thoroughly articulate SDT-based research trajectories to benefit HCI games scholarship
, and drive research towards more influential and cumulative outcomes. 
The following ideas are intended as starting points for more theory-informed work in HCI games research based in SDT, by linking games phenomena to SDT propositions, as well as developing theory-based resources for design practice. 


\subsection{Intrinsic Motivation and Functional Significance} 
We begin our examination of SDT-based research opportunities with \emph{cognitive evaluation theory} \cite[CET,][]{ryan2017cet}, 
which emphasises the relationships between basic need satisfaction, extrinsic rewards, and intrinsic motivation. Several of the reviewed HCI games papers commented on the potentially detrimental effects of extrinsic rewards on intrinsic motivation \citep[e.g.,][]{groening2019achievement,mekler2017towards}, and some preliminary tests of these effects have been conducted in recreational game contexts \citep{frommel2022daily,johnson2018rewards}. It is worth noting, however, that in-game rewards are typically integrated into a design such that they do not seem `extrinsic', and for this reason do not appear to undermine intrinsic motivation \citep{johnson2018rewards,phillips2018rewards}. What remains to be examined is the extent to which variations in the \emph{functional significance} \cite{ryan2017cet,thibault2017carrot} of in-game rewards can change their associated motivational outcomes. 
HCI games scholarship could, for example, investigate the 
conditions 
under which 
different rewards are more likely experienced as a means of controlling player behaviour, rather than informational feedback \citep[e.g.,][]{vanroy2019collecting}. Potential conditions include how directly rewards are linked to player performance \citep{cerasoli2014intrinsic,cerasoli2016performance}, whether they highlight players' effort versus ability \citep{fong2019negative}, or to what extent players are oriented towards task mastery \citep
{cerasoli2014mastery,fong2019negative} over meeting normative standards (e.g., relative to other players). 

To further %
extend (or refute) 
theoretical links across 
SDT concepts, 
CET could also be applied to 
explore 
known games phenomena. For instance, whereas CET considers control and autonomy conceptually distinct \citep{deci1985general} and posits that unexpected rewards do not affect intrinsic motivation \citep{deci1999meta}, recent HCI games work \citep{yin2022reward} theorised that randomised rewards (e.g., loot boxes) might thwart autonomy as players have limited control over what reward they receive. Similarly, studies on the motivational impact of `juicy' feedback \citep{hicks2019juicy,hicks2019understanding,tornqvist2021motivated} have yielded mixed results. Drawing from CET, it is possible that 
`juiciness' affects need satisfaction differently to the extent that visual embellishments confer 
informational feedback, 
thus augmenting perceived competence; or help amplify players' perceptions of their actions as originating from a more internal locus of causality \citep[see][]{andres2018super,Granqvist2018exaggeration}, supporting autonomy satisfaction. 

Collectible rewards \citep{hodent2016gamer,hodent2017gamer3,lewis-evans2017throwing}, such as the hundreds of items \citep{powerpyx2021assassin} in \emph{Assassin's Creed Valhalla} \citep{ubisoft2020assassin}, 
may also be fruitful sites of research, particularly for players who feel compelled to `100\%' the videogames they play. 
Further, there are opportunities 
to examine more exogenous design features common to live service games \citep[e.g., as in][]{cruz2017need,frommel2022daily} -- in particular, timers and other waiting mechanics \citep{keogh2018waiting,street2014delay}. Clearly these games are financially successful \citep{alha2016critical}, but it is worth considering the ways that the functional significance of these features can allow live service games to regularly obstruct player progress without damaging long-term motivation. 


Finally, CET could 
form the basis for a `generative theory of interaction' \citep{beaudouin-lafon2021generative} to help 
highlight aspects of interest 
and generate new ideas for game design \citep[akin to][]{deterding2015lens,schell2020art}. 
According to Beaudouin-Lafon et al., generative theories ``involve three successive lenses [...] The \emph{analytical} lens provides a description of current use and practice; the \emph{critical} lens assesses both the positive and negative aspects of a system [...] the \emph{constructive} lens inspires new ideas relative to the critique'' \citep[p.~3, emphasis in original]{beaudouin-lafon2021generative}. 
Taking functional significance as a grounding concept, for 
instance, the analytical lens may ask `in what ways does the game reward players' progress, performance, or mastery?'; 
the critical lens may question `to what extent is the reward relevant to the player's action(s)?'. Finally, the constructive lens helps consider alternate designs, for example, `how would 
removing reward(s) change players' interpretation of game events?'. 
Of course, these examples serve primarily illustrative purposes, and their utility for design is, at best, unclear. More principled efforts, however, 
could help integrate CET and design more tightly, and fruitfully 
extend 
theoretical tenets \citep[e.g., how players functionalise rewards,][]{phillips2018rewards,vanroy2019collecting}.

\subsection{Needs, Player Experience, and Behavioural Outcomes} 

\emph{Basic psychological needs theory} (BPNT) underpins much of SDT's conceptual apparatus. The notions of autonomy, competence, and relatedness are likewise prevalent in HCI games research, with seemingly wide-ranging consensus regarding their utility for industry. 
It has been suggested, for example, that 
they ``benefit game developers in offering them feedback that is not only game-oriented and actionable but also does not hinder the creative process of game development'' \citep[p.~3]{azadvar2018upeq}. Yet little is known about what makes these concepts 
actionable, or how they facilitate communication within the development team. %
In light of work linking 
need satisfaction 
to various industry success metrics (e.g., critical acclaim \cite{Johnson2014edge,ryan2006motivational}, purchase intention \citep{rigby2008sustaining,zhao2022exploring}, time spent playing \citep{azadvar2018upeq,johnson2016motivations}), it is also worth considering 
how (uses of) these concepts come to bear on designers' creative autonomy \cite[see][]{whitson2020what}. 
%

Similarly, as noted in section \ref{why}, there is a risk of need satisfaction becoming commodified \citep[see also][]{bennett2023does,soderman2021against}. Autonomy, in particular, has been linked to sustained engagement and future play intention \citep{rigby2008sustaining,ryan2006motivational}, as long as ``
[players] feel that they've chosen'' \citep{rigby2008sustaining}. In fact, although \emph{volition} is central to SDT's conceptualisation of autonomy \citep{koestner2021generative,vansteenkiste2020basic}, SDT games scholarship has to date primarily focused on \emph{in-game} choices \citep{rigby2008sustaining,rigby2011glued,rigby2017freedom}. Examining the ways players 
dis/re-engage with play 
\citep{hammad2021homecoming} or negotiate consent in games \citep{nguyen2020challenges} would help develop a fuller understanding of autonomy and its relation to player behaviour and wellbeing. 

With regards to need frustration, existing studies 
primarily capture psychosocial processes in daily life, 
and consistently focus on dysregulated play \citep[e.g.,][]{allen2018satisfaction,ballou2022people,mills2018exploring}. 
However, recent work 
linked \emph{in-game need frustration} to (self-reported) play behaviour \citep{ballou2022just,kosa2022need}, such as quitting intention. Preliminary evidence suggests that play-related outcomes 
occur as a function of the degree and frequency with which games unexpectedly frustrate player needs \citep{ballou2022just}. Likewise, further work connecting need frustration to other experiential concepts in HCI \citep[e.g., emotional challenge;][]{bopp2018odd} may effectively broaden our understanding of the ways that uncomfortable experiences can 
shape the player experience. The relevance of this work also finds support from game developers: Burnell \citep{burnell2012breaking}, for example, has 
suggested that designers can productively `break' players' basic needs to create experiences of heightened negative emotion that players ultimately value. 

Taking a more directly motivational perspective, Melhart \citep{melhart2018towards} has argued that players manage negative experiences or boring periods of videogame play by shifting between an intrinsic and extrinsic 
locus of causality. Others have similarly proposed that players may develop more internalised forms of extrinsic motivation to push through unpleasant sections of a game, depending on their prior enjoyment \citep[][p.~425]{tyack2017exploring}. While research in this area remains preliminary, further examination of the ways that motivation, need satisfaction and frustration, and other PX dimensions vary across different `sections' of a game could offer new insights into relationships between game design and player experience. 

Of course, absent cathartic payoff, protracted experiences of need frustration are simply unpleasant. When need satisfaction appears unattainable, people may turn to \emph{need substitutes} \citep[or `deficit motives';][]{ryan2000darker} such as security, self-esteem, or wealth \citep{sheldon2001satisfying}. It is arguably rare for videogame play to consistently frustrate players' needs over time; however, these circumstances may arise in multiplayer games with long match times (e.g., MOBAs). Studies of toxicity and other deviant behaviours may therefore benefit from an analytic frame incorporating need substitutes.

\subsection{Need Satisfaction, Wellbeing, and Immersion}
\label{needswell}
A fair number of HCI games studies 
concerned links between need satisfaction and wellbeing \citep[e.g.,][]{herodotou2014dispelling, poppelaars2018impact, tyack2021small, vella2015playing}, building on early SDT games research \citep{ryan2006motivational,przybylski2009having}. For this reason, relationships between in-game need satisfaction and short-term wellbeing are relatively well-established \citep{birk2013control, ryan2006motivational, tyack2020restorative}. 
%
Notably, the \emph{need density hypothesis} \citep{rigby2011glued,rigby2017time,ryan2017motivation} 
posits that videogames can impair player wellbeing precisely \emph{because} they satisfy basic needs more effectively, immediately, and predictably than other activities. However, besides one unpublished study \citep{rigby2011glued}, this claim is yet to be empirically supported. 
Findings from a recent cross-sectional study \citep{ballou2022people} remain inconclusive, and some work indicates that games and other screen media are comparably need-satisfying \citep{tothkiraly2019two}. Current explanations link immediacy to the mobility of smartphone games \citep{ryan2017motivation}, though 
public play can impede autonomy satisfaction 
\citep{deterding2016contextual}. 
Similarly, while games often ``provide goals and quests that can be rapidly completed with immediate positive feedback'' \citep[p.~519]{ryan2017motivation}, many popular games place limits on engagement (e.g., via weekly raid lockouts), and delay satisfaction using timers or energy mechanics \citep{keogh2018waiting,frommel2022daily,street2014delay}. %
In light of recent 
gaming trends, such as 
100-player Battle Royale games like \emph{Fortnite} \citep{epic2017fortnite} and highly random 
games like 
\emph{Hearthstone} \citep{blizzard2014hearthstone}, it is also difficult to argue that videogames present ``just world[s]'' \citep[p.~519]{ryan2017motivation} that satisfy needs in predictable ways. 
%
%
%
%
%
Questions remain, then, regarding the ways players seek out games for need-satisfying 
purposes, and under what conditions videogame play can be detrimental to wellbeing.

Need satisfaction \emph{in daily life}, for instance, appears a crucial factor \citep[e.g.,][]{allen2018satisfaction,milyavskaya2013psychological}. 
SDT games scholarship \citep{przybylski2009having,przybylski2019investigating} 
associated high levels of need satisfaction in daily life with more harmonious 
game engagement, whereas low need satisfaction in daily life was linked to more obsessive play. More recent investigations, however, 
found \emph{need frustration in daily life} to be a stronger predictor of obsessive engagement 
\citep{formosa2022need,tothkiraly2019two}. Evidence is similarly mixed 
regarding the extent to which need satisfaction or need frustration in daily life 
predict in-game need satisfaction 
\citep{ballou2022people,bender2020internet,formosa2022need,johnson2022unsatisfied}. Interpretation of results is further complicated by these studies' deployment of a need frustration scale \citep[i.e.,][]{chen2015basic} whose validity was recently questioned \citep{murphy2022basic}. As such, there is a need to identify conditions 
that help account for these mixed findings, alongside potential methodological confounds. 
Ballou et al.~\citep{ballou2022people}, for example, theorised that play-related wellbeing benefits 
vary to the extent that players 
devalue 
the `genuineness' of the experienced need satisfaction (e.g., ``it's only a game''). 
Similarly, lack of autonomy satisfaction in daily life might contribute to amotivated play \citep[i.e., because ``there's nothing else to do'',][p.~8]{turkay2022self}, %
which 
may undermine 
wellbeing benefits 
derived from play 
\citep{vuorre2022time}. 

Recent large-scale studies \citep{johannes2021video,vuorre2022time} 
indicate that need satisfaction, intrinsic motivation, and extrinsic motivation\footnote{Measured via an 
undisclosed variant of the PENS.} independently predict wellbeing (as operationalised via positive and negative affect, as well as life satisfaction), 
and that wellbeing is related to subsequent playing motivation \citep{vuorre2022time}. These results suggest a reciprocal relationship 
\emph{over time}, but explanatory accounts are missing at present. Considering motivation, needs and wellbeing at different levels of generality \citep[as suggested in][see also section \ref{person}]{ballou2023bangmodel} could help clarify some of these processes. 
Moreover, although research has shown that games 
can restore wellbeing after a negative event \citep{tyack2020restorative, tyack2021small}, 
an equally interesting question is whether the benefits of videogame play can also \emph{protect} players against psychological deficits caused by a subsequent negative event. Lastly, there is a wealth of literature outside SDT on videogames and coping behaviour \citep[e.g.,][]{iacovides2019role, wolfers2020media} that could be fruitfully incorporated into further studies; for example, by examining emotion regulation \citep[as in][]{caro2021gaming} alongside more commonly used SDT wellbeing measures such as vitality. 


%
%

Finally, games research is well-positioned to put wellbeing itself under greater scrutiny: Mindful awareness is increasingly essential to SDT's conceptualisation of wellbeing \citep{brown2003benefits, rigby2014mindfulness, ryan2017basic} -- yet, as previously noted, scholarship from SDT scholars and others has indicated that mindful awareness and immersion are antithetical \citep{dane2011paying, sheldon2015experiential}. The relationship between immersion and mindful awareness could be productively complicated by these concepts' concurrent examination in the context of videogame play. 
Indeed, SDT's view of immersion is less frequently deployed than other perspectives in HCI games research \citep[e.g.,][]{brown2004grounded, jennett2008measuring}, potentially because immersion's links to other SDT concepts are largely indeterminate. Specifically, it is unclear whether immersion is more appropriately used as a moderator \citep[as in][]{przybylski2010motivational,przybylski2012ideal}, an independent variable, or a meaningful dependent variable in its own right \citep[as in][]{przybylski2009motivating, ryan2006motivational}. 
For PENS immersion to usefully inform SDT-based models of player experience, further work is needed to theorise 
and empirically test its relationships with other core SDT concepts, and immersion's importance for player experience. Recent work by Mella and colleagues \citep{mella2023gaming}, for instance, suggests that immersion \citep[as measured by][not the PENS]{jennett2008measuring} contributes to post-work relaxation, as well as recovery of a sense of mastery and control -- which overlap conceptually with competence and autonomy satisfaction.


\subsection{Extrinsic Motivation and Regulatory Styles}
HCI games research involving extrinsic motivation has typically concerned the ways that regulatory styles differentially influence play-related outcomes. 
In this way, \emph{organismic integration theory}'s \citep[OIT,][]{ryan2017organismic} conceptual breadth has been used to productively investigate a range of topics, such as dysregulated play \citep{mills2018gaming,mills2020self}, passion \citep{wang2011understanding}, persistence \citep{neys2014exploring}, and 
engagement rewards \citep{frommel2022daily}. %
Future work may consider to what extent 
OIT-based findings on (professional) sports \citep[e.g.,][]{cece2020self,li2013burnout,ntoumanis2001empirical} generalise to esports, or how regulatory styles relate to players' experience of (accepting) failure \citep{frommel2021struggle,sheldon2007obligations}. 

Further, SDT-based gamification research often frames intrinsic motivation as the primary driver of long-term engagement \citep[e.g.,][]{lessel2019enable,mekler2017towards,sailer2017how} for its apparently self-maintaining quality, whereas extrinsic motivation is commonly reduced to external regulation \citep{cermak2022should}, and considered to require greater maintenance \citep[e.g., via reward schedules;][]{landers2019greatest}. Autonomous forms of extrinsic motivation  \citep{burton2006differential}, however, may hold substantial value for gamification and applied games research \citep[e.g.,][]{aufheimer2023examination,birk2016motivational,birk2018combating,cheng2019gamification}. For example, scholarship in other domains \citep[e.g.,][]{mankad2014motivational} has identified links between autonomous extrinsic motivation and behavioural engagement -- an essential quality for games or game-adjacent systems whose success (or profit) is predicated on long-term player retention. Further work in this area should, therefore, consider \emph{autonomous} forms of motivation, rather than intrinsic motivation alone. 

%


Recent SDT-informed work has also employed statistical clustering methods \citep[e.g.,][]{fernet2020self,howard2018using,howard2021longitudinal} to identify motivational profiles, that is, distinct patterns of \emph{co-occurring regulatory styles}. 
A study on recreational MOBA play \citep{bruhlmann2020motivational}, for instance, 
observed that already 
slightly elevated levels of amotivation and controlled motivation impede enjoyment and need satisfaction, even when intrinsic motivation is high overall. 
Questions remain, then, regarding the ways motivational regulations \emph{shift} over time \citep{gunnell2014goal,wasserkampf2016organismic}, and how this relates to psychosocial and behavioural outcomes (e.g., churn, wellbeing). According to OIT, more need satisfaction is bound to (eventually) foster higher levels of autonomous motivation (i.e., identified, integrated, intrinsic). 
However, it is not clear whether shifts towards autonomous motivation are determined by the frequency or degree of (in-game) need satisfaction, and how long it takes for motivational changes to (demonstrably) manifest. 

On this note, there is much promise in investigating the ways games 
facilitate \emph{internalisation} \citep{deci1994facilitating,ryan2017self}, the process through which behaviours become organised with other facets of the self. 
Drawing from SDT scholarship on parenting and education \citep[e.g.,][]{ryan2017parenting,vansteenkiste2018fostering}, Tyack and Wyeth \citep{tyack2017exploring} propose that videogames encourage internalisation by communicating care and autonomy support towards the player, as well as the provision of structures that facilitate experiences of competence, ``which can act as an ongoing support in moments of both insecurity and growth.” \citep[p.~347]{ryan2017parenting}. Similarly, Rigby suggested that a heroic frame amplifies the \emph{personal relevance} of game quests \citep{rigby2009virtual,rigby2011glued}, a known predictor of internalisation  \citep{deci1994facilitating,vansteenkiste2018fostering}. 
Beyond the player-game relation, factors such as parental communication style towards videogame play \citep{bradt2023does,vanpetegem2019parents} are also worth considering. Controlling (versus autonomy-supportive) communication, for example, is more likely to foster partial internalisation \citep{ryan2017parenting} and introjected regulation
, which might in turn manifest in obsessive passion \citep{vallerand2003passions,przybylski2009having} and devaluation of play \citep{ballou2022people}.  
Finally, research is needed to identify conditions that allow \emph{amotivation} to supplant initially motivated play \citep{ntoumanis2004idiographic}. Potential candidates include lack of autonomy satisfaction in daily life \citep{turkay2022self} and repeated 
in-game need frustration \citep{ballou2022just,tyack2021small}. 

Ultimately, cumulative OIT-based research could inform investigations into how regulatory styles determine the player experience at different levels of generality 
\citep[][see next section]{blanchard2007reciprocal,vallerand1997toward}, for example, the extent to which externally regulated pursuit of an individual 
sidequest \citep[e.g., `Blitzball' in \emph{Final Fantasy X,}][]{ffx2001} might infringe on players' overall intrinsic motivation to play a particular game \citep{tyack2017exploring}, or even an entire genre. 

\subsection{Person- and Context-Level Factors}
\label{person}
Much SDT-based games research currently operates at an \emph{episodic} (or `situational' \citep{vallerand2000deci}) level of analysis; that is, with respect to a single session of interaction with a game. 
In contrast, \emph{causality orientation theory} \citep[COT,][]{ryan2017causality} and \emph{goal content theory} \citep[GCT,][]{ryan2017goal} are individual-difference theories -- and as such, they draw attention to broader units of analysis: COT focuses on person-level factors, whereas GCT corresponds to particular social contexts (or domains). Being explicitly based in SDT, these mini-theories 
provide more concepts and propositions to work with, than models that are currently popular in HCI games research. The Hexad \citep{tondello2016gamification}, for example, claims to have a basis in SDT, but its model is inconsistent with the theory in many ways \citep[p.~7]{tyack2020self}. SDT frameworks such as the hierarchical model of motivation \citep{blanchard2007reciprocal,vallerand1997toward} 
demonstrate the importance of conducting research at the primary level of analysis -- personal, contextual, or episodic -- while remaining attentive to indirect influences from factors at other levels \citep[also see][]{milyavskaya2013psychological}. Indeed, a small number of studies \citep{melhart2018towards, vanhoudt2020disambiguating} have already productively applied the model to produce more granular perspectives on player experience. In this sense, the hierarchical model of motivation functions similarly to the more recent METUX model \citep{peters2018designing}, in that the latter also proposes a number of `spheres of existence' in which technology can operate to varying degrees. 



HCI games research that incorporates player typologies and modelling may consider applying COT as a more theory-based alternative to existing frameworks. In gamification research, for example, causality orientations can be readily implemented as moderators \citep[e.g., as in][]{mekler2017towards}. Although goals can operate at higher levels (e.g., life goals), the most productive opportunities for applying GCT to games research are arguably situated in contexts of play \citep[e.g.,][]{fernandez2023last}. For example, as in-game transactions and possessions become increasingly prevalent in videogame design \citep{alha2016critical, nieborg2015crushing}, particularly cosmetic items, goals such as (visible) wealth, status, and materialism may become more salient \citep{carter2020situating}. This assetisation of triple-A videogames \citep{bernevega2021industry} also intersects with livestreaming cultures, as streamers' in-game possessions can become normalised through visibility, and potentially more desirable as a result. Motivations towards purchasing and owning particular virtual objects \citep{toups2016collecting,zhao2022exploring}, particularly those subject to changes in game balance, can be complex. Players may feel personally attached to a weaker game character, for example, or expect balance changes in the future. Understanding these motivations requires analysis with a similarly robust theoretical basis, for instance, in terms of how goals, motivational regulation and need satisfaction influence each other over time \citep[as in][]{gunnell2014goal}. 



Notably, power, dominance, and performance goals -- which particularly feature in competitive multiplayer play -- have been identified as extrinsic within GCT \citep{ryan2017goal}. Further study into these goals may help us better understand negative outcomes of such games, such as player burnout, ill-being, and 
deviant behaviour \citep[or `toxicity';][]{kou2020toxic}. Recent work indicates, for example, that collegiate esports players normalise (and themselves propagate) deviant behaviour \citep{turkay2020see}, and GCT-based research may offer further insights into the links between performance goals, deviance, and coping mechanisms. These goals also represent a potential point of convergence with Motive Disposition Theory \citep[MDT;][]{mcclelland1985motives}, which has recently seen greater attention in HCI games research \citep[e.g.,][]{poeller2018implicit, poeller2021prepare} -- in particular, MDT's achievement and power motives are comparable to performance and power goals in SDT. In this way, goal contents could be applied 
to differentiate players (e.g., with social vs. performance goal contents), or better understand the experiences and behaviours of players with a particular goal.

\subsection{Social Play}
\label{social}
SDT -- and \emph{relationship motivation theory} (RMT) in particular -- emphasises the importance of reciprocal need support, autonomous motivation, and non-contingent care for others for high-quality relationships \citep{deci2014autonomy,ryan2017relationships}. These concepts have to date been almost completely overlooked in HCI games scholarship; however, social play may represent a particularly salient site for RMT-based research. 
For instance, research into couples' cooperative play \citep[e.g.,][]{butt2016girlfriend,ratan2015stand} indicates that women's displays of skill may be met with disapproval by their (male) partners; moreover, they are often pressured into supporting roles. In RMT terms, these works showcase instances of 
\emph{conditional regard} \citep{kanat-maymon2016controlled, roth2012costs} towards partners who do not accede to expectations \citep[as in][p.~49-50]{butt2016girlfriend}. Further, deviant behaviour occurs among friends (and strangers) who play games together, and RMT offers both a means of understanding these situations, and concepts \citep[e.g., mutual need support,][]{deci2006benefits,itzchakov2023connection} around which game design could be organised to promote 
positive in-game socialisation. Recent work \citep{frommel2023perceived}, for instance, found that social capital partially mediates the impact of perceived toxicity on relatedness satisfaction. 
Building community infrastructures that promote 
mutual need support 
could hence 
foster 
social %
bonding, and perhaps even counter (perceptions of) toxicity. 

RMT concepts could also 
inform 
intermediate-level knowledge \citep{dalsgaard2014between,stolterman2010concept} to analyse existing artefacts and inspire new designs. With regards to mutual need support, for instance, generative questions \citep[see][]{beaudouin-lafon2021generative} might pertain to
`what opportunities are there for players to support others' needs?' (analytical) and `
does the game allow players to %
acknowledge 
others' need support?' (critical). 
One of the endings in \emph{NieR: Automata} \citep{nier2017}, for example, places the player in a nigh-insurmountable encounter, unless they accept help from other players. The player later has the choice to help other (unknown) players 
-- at the cost of their own save file \citep{gallagher2018memory}. 
If the player accepts, they are not accorded any mechanical benefits, additional content, or obvious acknowledgment from others. 
Following from this example, a constructive question \citep{beaudouin-lafon2021generative} 
could therefore address `how difficult should it be for players to accept and provide need support?'. Again, these questions are but preliminary examples whose utility remains to be evaluated; but they highlight how theory translation efforts grounded in RMT could open up new avenues for design and theory development. 

Some work has also considered \emph{para}social play behaviours towards game characters \citep{azadvar2018upeq,bopp2019exploring,tyack2017exploring}, but SDT-based empirical research is currently lacking -- although in early SDT scholarship, researchers were purportedly ``intrigued by how needs for relatedness may be met by `computer generated' personalities and artificial intelligence'' \citep[][p.~350]{ryan2006motivational}. Players report strong feelings of relatedness towards game characters \citep{sutherland2020affect} and livestreaming `personalities' \citep{carter2020situating, lu2021kawaii}, even though these parasocial relationships necessarily cannot offer the requisite mutuality that RMT espouses. These relationships may therefore represent an opportunity to extend the theory to accommodate relationships that primarily offer an illusion of care and need support. Emotional exploration \citep{cole2022emotional} and perspective-taking \citep{grasse2022using} were recently 
proposed as relatedness-supporting devices in single-player games, for example. 

It is also worth considering relatedness 
at the 
\emph{group} level 
\citep[e.g.,][]{amiot2012can,kachanoff2020free, kachanoff2019chains}. 
Research on the motivational antecedents of deviant behaviour in team sports \citep{mallia2019athletes,ntoumanis2009morality} 
could inform work aimed at understanding 
the emergence of toxicity in esports. 
More broadly, Tyack and Wyeth \citep{tyack2017exploring} have 
theorised relatedness at the \emph{cultural} level (e.g., hegemonic `gamer' culture, a particular fandom), drawing from Bourdieu's notion of group \emph{habitus} \citep{bourdieu1990logic} -- i.e., a shared set of practices, experiences, and cultural references that do not require 
explicit co-ordination with others. 
Specifically, the authors posit that players experience relatedness 
to the extent that they perceive themselves as belonging to a cultural group that 
accepts and validates an aspect of their `true self'. Conversely, relatedness is threatened by 
exposure to 
divergent habitus \citep[e.g., `walking simulators' that do not accede to players' expectations of what a game `ought to be',][]{grabarczyk2016walking,muscat2020mechanics}. In line with
SDT research on inter-group discrimination \citep{amiot2012can}, 
hostility directed towards players and developers of divergent habitus may thus be an attempt to compensate for need frustration via conforming to in-group habitus. 


Lastly, while early HCI games research often assumed that social play was universally better than solitary play \citep[e.g.,][]{cairns2013who}, more recent work has demonstrated that player autonomy can be negatively affected by other people's presence in the game, or physical proximity \citep{deterding2016contextual,vella2015playing}. There remain compelling opportunities to investigate the ways that different degrees of 
co-presence (e.g., co-located play, voice interaction \citep{carter2015player,poeller2023tekken}) can influence players' quality of motivation and sense of autonomy towards play, and relatedness to the co-present other(s).

\section{Implications for Theory in HCI}
\label{hci}

While the present work focused on SDT and games, our findings speak to wider considerations of theory in HCI. 
Indeed, many of the issues we identified 
are not unique to SDT, 
but 
resemble those documented elsewhere in the HCI literature: Cherrypicking of a few concepts over leveraging theoretical tenets \citep{hekler2013,velt2017trajectories}, tenuous links between theory and design \citep{beck2016examining,oulasvirta2022counterfactual}, theoretical misconceptions \citep{bardzell2009interaction,hekler2013}, and few efforts to extend or develop theory \citep{hornbaek2018commentary,velt2017trajectories}. In this sense, our work aligns with wider concerns around the 
paucity of 
theory-based research in HCI \citep{beaudouin-lafon2021generative,hornbaek2018commentary,kostakos2015big,oulasvirta2022counterfactual}. 

In our previous review \citep{tyack2020self}, we lamented perfunctory uses of SDT in HCI games research, but without considering reasons for this, beyond a presumed unwillingness to engage with the theory more deeply. 
%
%
The present work paints a more complex picture: In examining SDT from the perspectives of HCI games research, industry practitioners, as well as SDT scholarship, we could clarify how the theory likely came to attain widespread popularity; not so much for its capacity to drive research or inform design, but largely for methodological and rhetorical purposes, which 
slotted into researchers' and game developers' existing practices and concerns. Further, our findings lend little credence to the claim that iterative design practices or anti-scientific sentiments inhibit theory use in HCI \citep{oulasvirta2022counterfactual}. Instead, we have 
traced 
HCI games researchers' limited 
theoretical engagement 
(in part) to issues inherited from SDT games scholarship, as well as the 
conditions 
that contributed to the theory's uptake in games research and industry in the first place. 

How to support more productive theory use in HCI? 
Our findings 
are specific 
to SDT in HCI games research
, 
and hence do not readily transfer to other theories or HCI at large. 
%
Moreover, the plurality of theories and theory purposes in HCI \citep{bederson2003theories,rogers2012hci} 
fundamentally resists recommendations 
that aim to cut across the field as a whole. 
Our work nevertheless points towards 
some 
initiatives 
for cultivating more \emph{intentional} practices of theory use that apply 
beyond SDT: 

First, 
we observed many instances in our corpus where SDT was referenced but ultimately inconsequential to the research. While it seems unlikely that these authors engaged in perfunctory citations on purpose, we suspect that few have actively considered the role of the theory for their work. 
Beck and Stolterman's models of theory use \citep{beck2016examining} 
could serve as a high-level epistemic tool to  
think about how theory is deployed 
in relation to one's research questions, method, and findings. For example, whether theory serves as a `contextual tool' \citep[p.~131]{beck2016examining} or `shaping tool' \citep[p.~132]{beck2016examining}, i.e., to situate one's research question or shape it through theoretical considerations. 
Though these models oversimplify the ways theory can inform research (e.g., they do not distinguish between design and empirical research), they may nevertheless support HCI scholars in engaging theory more deliberately. For instance, to facilitate more transparent documentation of theory use, or reflect on how 
the research 
might differ if it were shaped by an alternate theory. 
%
%
%

Second, Beck and Stolterman 
\citep{beck2016examining} emphasise 
`\emph{talking back}' to theory. 
As noted, 
many works in our academic corpus appeared reluctant to challenge SDT, even when results diverged, 
which conveys the impression that the theory is 
`complete' and sacrosanct. `Explicit talkback' to theory \citep[p.~137]{beck2016examining} -- i.e., comparing research findings with theoretical assumptions, discussing the utility and limitations of a particular theory for one's work, or suggesting 
revisions -- would help resolve and pre-empt such misconceptions, and %
highlight avenues for 
theory development. 

Third, we commented on the circumstances that helped cement SDT's status as 
the seeming `default' theory in HCI games research \citep[p.~261]{poeller2022self}. Our personal experiences and conversations with colleagues attest to a worrying consequence thereof: Reviewers sometimes critique submissions for \emph{not} drawing on SDT, and authors who employ alternate frameworks are challenged to justify their theoretical choices \emph{relative} to SDT. Frustratingly, these critiques appear rarely grounded in cogent and topical arguments, but instead invoke SDT's popularity in HCI 
-- often by recourse to our previous work \citep{tyack2020self}. 
We stress that this kind of theoretical gatekeeping is not only incorrect and unhelpful, but ultimately impedes progress in the field by reinforcing the dominance of a single theory \citep[see][on the merits of theoretical pluralism]{chang2012water}. We instead recommend that reviewers more carefully consider and, where pertinent, ask for clarification why a particular theory was deployed, rather than why a (dominant) theory was not used. For authors, we nonetheless advise engaging with concepts and theories that are (more) well-known within a particular research community, even when these did not inform the work -- not to pay theoretical lip service, but to `talk back' and explain 
why the dominant perspective was unsuited for that line of inquiry. 

%
Finally, HCI 
stands to benefit from greater attention to %
%
the normative implications of theory \citep{gergen1978toward,ghoshal2005bad,zoshak2021beyond} 
%
-- 
especially with regards to perspectives that (as with SDT in games\footnote{See also Soderman \citep{soderman2021against} for a related discussion of flow theory.}) dominate current academic and industry discourse. Neoliberal tenets, for example, 
inhere within many strands and theories of 
psychology \citep[incl. SDT,][]{adams2019psychology,van2023critiques}. %
In applying these theories to technology design and evaluation \citep[e.g., for behaviour change,][]{consolvo2009theory,villalobos-zuniga2020apps}, HCI scholarship 
reinforces (sometimes unwittingly) the same ideological underpinnings \citep[e.g., over-emphasising individual responsibility,][]{brynjarsdottir2012sustainably,kou2019turn}. 
%
%
%
%
%
Of course, the normative implications of a theory need not (invariably) preclude its merits for research and design. 
Rendering theoretical presuppositions more apparent, however, would help denaturalise 
taken-for-granted assumptions 
and strengthen 
researcher reflexivity. The practice of diagramming
`field theories' \citep{ploderer2021diagramming} seems particularly suited to this end. Although originally devised as a pedagogical tool for 
developing intermediary 
design knowledge, 
this approach could be leveraged 
to `draw' out and compare implicit theoretical assumptions (e.g., what understanding of `good' interaction is inherent to a particular theory?). Given the strong emphasis on field research, this approach could also help call attention to 
the ways that (realising) a theory's normative implications 
might clash with or
infringe on local concerns 
\citep[see also][]{adams2019psychology,ghoshal2005bad,tomlinson2013insubordinate}. 

\section{Limitations and Open Questions}
\label{limitations}

We have reviewed SDT's application to games from three perspectives -- within SDT scholarship, in HCI games research, and at GDC. The broad scope of our analysis inevitably comes with a number of caveats and leaves open questions for future work: First, in face of the impossibility to further consult with the first author -- circumstances whose infinite shittiness defies all description -- the second author decided not to extend the literature review to papers published after 2021. Informal examination of more recent HCI games papers suggests, however, that the patterns and issues we identified in the present work largely persist in current scholarship.

Second, the 
publications in our HCI games corpus were described only briefly. While this was necessary to avoid bloating an already extensive paper, much more could be said about the reviewed works. Examining to what extent supporting and reiterating citations centre 
distinct aspects of SDT, for instance, could help further characterise the rhetorical 
power of the theory \citep{beck2016examining,halverson2002activity} relative to other motivational frameworks \citep{poeller2022self}. Moreover, 
our coding schema categorises theory use with regards to the \emph{citing authors' claims}, even when they conflict with SDT. Although the analytic and generative categories represent more extensive applications of SDT, our results show that this did not preclude theoretical misunderstandings. 
We note that this tension is inherent to the citation typology \citep{girouard2019reality} we built our coding schema on. Further granularity may render the theory use categorisation more insightful, for instance, to assess how faithfully a study's hypotheses reflect SDT mini-theories and their propositions, or to detail how theoretical tenets inform study design and game stimuli selection \citep[see][]{tyack2018video,vornhagen2020statistical}.

Third, 
the venue selection for the academic literature review was limited, though we believe it provides an adequate snapshot of HCI games scholarship. 
It is worth noting, however, that the inclusion of CHB -- whose status as a core HCI venue is disputable\footnote{Unlike other venues included in our analysis, CHB is neither listed in the \href{http://portal.core.edu.au/jnl-ranks/}{CORE journal rankings} nor among the top HCI venues as per \href{https://scholar.google.com/citations?view_op=top_venues&hl=en&vq=eng_humancomputerinteraction}{Google Scholar}.} -- actually \emph{increased} the number of analytic theory use instances in our corpus. Had we omitted CHB from our analysis, results would have skewed slightly further towards perfunctory applications of SDT (i.e., descriptive n = 107, 59.12\%; analytic n = 57, 31.49\%; generative n = 17, 9.39\%). 

Of course, our analysis could be fruitfully extended to other 
HCI venues, such as \emph{Human-Computer Interaction} \citep[e.g.,][]{deterding2015lens}, \emph{ACM Interaction Design and Children} \citep[e.g.,][]{mcewan2020puppy}, 
\emph{Behaviour \& Information Technology} \citep[e.g.,][]{turkay2022self}, or the recently launched \emph{ACM Games} journal. Further, investigating SDT-based games research in other fields, such as media psychology \citep[e.g.,][]{bender2020internet,tamborini2010defining,valkenburg2013developing}, developmental psychology \citep[e.g.,][]{bradt2023does,vanpetegem2019parents}, or game studies \citep[e.g.,][]{conway2017sein}, could highlight additional avenues for theory-informed HCI games research. 

Moreover, 
SDT is an increasingly popular theoretical framework in HCI well beyond games \citep{ballou2022self}. Examining its applications for design and evaluation \citep[e.g.,][]{lallemand2014relevant,xi2021designing,zhao2023older}, or when theorising about the user experience \citep[e.g.,][]{hassenzahl2010needs,hornbaek2017technology,peters2018designing}, could prove insightful to ascertain the extent to which similar issues 
obtain in HCI-adjacent SDT research. 
Research based on the Technology Acceptance Model \citep[e.g.,][]{rezvani2017motivating}, for example, often draws from SDT's distinction of intrinsic and extrinsic motivation, yet has long overlooked the role of need satisfaction \citep{hornbaek2017technology}. Similarly, the METUX model of technology use \citep[][co-authored by Ryan]{peters2018designing} ignores existing SDT scholarship on the hierarchical model of motivation \citep[e.g.,][]{vallerand1997toward}, despite considerable conceptual overlap. 

That said, how HCI (games) scholars \emph{cite} SDT 
might not fully reflect the ways they actually \emph{apply} SDT in their work. 
Observational accounts of the ways that HCI scholars 
employ theory in their research practice 
could 
help identify theory uses currently not captured in existing categorisations \citep{bardzell2016commentary}, as well as 
explicate why (a particular) theory can be challenging to put into use \citep{halverson2002activity}. 
Our analysis further demonstrates that citation counts \citep{girouard2019reality,rogers2012hci}, or other traditional metrics \citep[e.g.,][]{kostakos2015big}, are unlikely to fully capture the influence of a theory. 
As such, we propose greater attention towards the \emph{indirect} ways that 
theory informs HCI scholarship and design practice. 
For instance, with regards to 
a theory's rhetorical function in industry practice \citep{clemmensen2005community,gray2014reprio}, or the ways theories are taught in HCI programmes \citep[e.g.,][]{girouard2019reality,ploderer2021diagramming}. 


Speaking to industry practice, our analysis did not take into account more recent GDC talks and there are 
presentations we missed because 
no slides were available \citep[e.g.,][]{hecker2010achievements,rigby2017freedom,vandenberghe2016engines}. Moreover, as a highly exclusive event that costs upwards of \$1000 USD to attend, GDC cannot be considered wholly representative of the games industry
, but its influence as a cornerstone of knowledge sharing and networking is widely acknowledged \citep{engstrom2019gdc,francis2021new}. GDC presentations also provide less material to review than a paper, which makes assessing theory use more difficult. It would be worth investigating the ways that practitioners actually apply SDT (or other theories) \emph{in situ}, relative to their discussions of theory use in game convention presentations. Such inquiries \citep[as in][]{velt2020translations} could help explicate the processes 
involved in translating theory into a useful design resource. 
Although we acknowledge that gaining access to game development environments can be challenging, prior ethnographic work on game production \citep[e.g.,][]{banks2009expertise, bulut2020precarious, whitson2017voodoo, whitson2020what} indicates that 
access is often permitted under some conditions (e.g., company anonymity). 

Lastly, it would be 
interesting to investigate how 
game developers adapt existing theory into their own folk theories. Ubisoft's engines of play framework \citep{vandenberghe2016engines}, for instance, draws both from SDT and personality psychology. 
More critically, said framework (and the GDC talks we reviewed) exemplifies how 
theory is sometimes co-opted by major game studios as a yardstick for `good' PX \citep{soderman2021against}; concurrently, general player experience models, such as the PENS, largely reflect standards of the historically dominant triple-A industry \citep{ruberg2015no,tyack2021offpeak}. In this way, theory reinforces the hegemony of the formal games industry, granting 
scientific 
legitimation to further ``determine(d) how videogames would be evaluated [by players\footnote{Rigby \citep{rigby2017freedom}, for example, justified the backlash against 
\emph{No Man's Sky} \citep{klepek2016inside} -- developed by a 
team of six employees, on average 
\citep{crecente2019disastrous,keogh2023triple} -- in terms of the game not providing players with enough
``dense volitional content to consume and be satisfied by'' \citep{rigby2017freedom}, comparing it to titles such as \emph{Grand Theft Auto V} and \emph{Witcher 3}, which were produced by studios with considerable financial backing and staff numbering in the hundreds.}, the market ...] %
in such a way that only the formal industry would have the resources and ability to develop and distribute videogames that would be evaluated 
as being of good quality'' \citep[p.~22]{keogh2019aggressively}. 
Greater attention to the ways SDT and other theories potentially impinge on more \emph{informal} game development practices \citep[][e.g., of independent gamemakers, hobbyists]{keogh2019aggressively,keogh2023videogame} is therefore needed, lest they be treated as 
peripheral sites for theory application. 
To this end, initiatives to translate theory \emph{in collaboration with} game developers 
offer a key opportunity for HCI games researchers to tangibly demonstrate the conceptual and practical utility of their work.  



\section{Conclusion}
Much of HCI games research has relied on self-determination theory (SDT) for concepts and measures that structure investigation of player motivation, experience, and wellbeing. This review has taken a more expansive view of the ways that SDT obtains in games research and development practice -- examining SDT scholarship on games, a wider corpus of SDT-based HCI games research, and perspectives from game development practitioners. In doing so, we have achieved four core aims. First, we have extended our prior review of CHI and CHI PLAY papers \citep{tyack2020self} to a number of other venues, observing stronger engagement with SDT tenets -- and yet we also noted 
few attempts to extend or meaningfully critique aspects of the theory, even when results are inconsistent with SDT. Second, we have identified issues with SDT games scholarship -- theoretical inconsistencies, unfounded claims, and limited engagement with other SDT concepts and mini-theories -- which may have contributed to perfunctory and tenuous applications of the theory in HCI games research. 
Third, our analysis of GDC presentations identified a common understanding of SDT concepts among game development practitioners, and a willingness to adapt or critique the theory that was not observed in our academic corpus. However, discussion of HCI games research findings was essentially non-existent. Finally, we discussed the conditions that facilitated SDT's uptake in academic and industry games discourse, and how these may have shaped -- and limited -- the ways HCI games researchers apply the theory at present.

The implication of our work is not that SDT 
is `imperfect' or `incomplete' -- theory development is an ongoing, iterative, and collective endeavour. 
Rather, 
the central issue is that 
(1) HCI games researchers have 
\emph{unquestioningly} adopted a theory 
that 
does not meet the conceptual and empirical standards its creators espouse, and (2) that \emph{continued 
neglect} to more deeply engage with SDT tenets \emph{perpetuates} perfunctory uses of the theory, which ultimately  impedes our capacity to develop HCI games knowledge and contribute to game design practice. 
Instead, HCI games research stands to benefit from adopting more intentional practices of theory use, by leveraging and developing SDT propositions to reason about games and play, and talk back to theory. 
To 
support these efforts, we have identified opportunities for HCI games scholars to bridge theory and practice, through 
engagement with game design practitioners and theory translation work. 
We have also proposed new avenues for SDT-based HCI games research in areas of un(der)developed theory application. 
These initiatives collectively underscore SDT's ongoing utility for games scholarship -- if HCI researchers can more effectively apply it. 

\section{Author Contribution Statement}
AT and EDM jointly developed the idea and direction of the paper. AT conducted the search of academic literature and GDC presentations, analysed all of the GDC corpus and 2/3 of the academic corpus, and contributed the majority of the writing. EDM contributed to the review of academic work and writing, replicated the literature search procedure in July 2022, and edited the final manuscript.

\begin{acks}
Special thanks to Dan Bennett, Feng Feng, Olga Iarygina, Cody Phillips, Susanne Poeller, Raquel Robinson, and Jan Vornhagen for providing feedback on earlier versions of this manuscript.

Funded by the European Union (ERC, THEORYCRAFT, 101043198). Views and opinions expressed are however those of the authors only and do not necessarily reflect those of the European Union or the European Research Council Executive Agency. Neither the European Union nor the granting authority can be held responsible for them. 
\end{acks}


%
\nocite{Alankus2012reducing,Alexandrovsky2019game,Altmeyer2019gamified,Andrade2019blind,Andres2018super,Balaam2011motivating,Baldwin2016crowd,Barathi2018feedforward,Bian2018weak,birk2013control,Birk2015self,Birk2016fostering,Birk2016motivational,Birk2016social,Birk2017age,Birk2018combating,Bopp2016negative,bopp2018odd,bopp2019exploring,Bowey2017dirty,Bowey2015leaderboards,Bowey2017stories,Cairns2014controllers,Cechanowicz2014improving,Charleer2018dashboards,Cmentowski2019outstanding,constant2019dynamic,Denisova2016convergence,Denisova2019rewarding,Denny2018empirical,Depping2018friendship,Depping2016disclosing,Depping2016trust,Depping2017cooperation,Depping2017happening,deterding2016contextual,Dodero2014tangible,Dominguez2016mimesis,Emmerich2016influence,Emmerich2017impact,Feger2019gamification,Fitz-Walter2014achievements,Gerling2014embodied,Gerling2014balancing,Gerling2015long,Granqvist2018exaggeration,Guckelsberger2017predicting,Gutwin2016helping,Gutwin2016peak,Hall2014instructional,Hallifax2019factors,Harteveld2015goal,Harris2019asymmetry,Hicks2015twitter,hicks2019juicy,Ioannou2019virtual,Johanson2016scaffolding,johanson2019pause,Johnson2014edge,Johnson2015cooperative,Johnson2015base,Kajastila2016augmented,Kao2018badges,Kaos2019social,Kappen2017gamification,Klarkowski2016operationalising,Kou2018playing,Krekhov2017self,Krekhov2019beyond,Kumari2019uncertainty,lehtonen2019movement,Lessel2019crowdjump,Lessel2019enable,Li2019book,Lomas2013optimizing,Lomas2017difficulty,Long2018latency,Long2019latency,MarquezSegura2016playification,McEwan2014natural,Morrison2009bees,Orji2018personalizing,Palacin-Silva2018gamification,Petralito2017good,Pfau2018deep,phillips2018rewards,Pillias2014tangible,poeller2018implicit,Poretski2019virtual,Reinschluessel2016using,Richards2014designing,Roohi2018intrinsic,Rooksby2015ball,Sarkar2019transforming,Shaer2017gaming,Smeddinck2016difficulty,Soroush2018investigating,Steinberger2017gamified,Tondello2016gamification,Tondello2017framework,Tondello2019player,Tyack2016appeal,VandenAbeele2015game,vella2015playing,Vella2017agency,Vella2018applied,Vicencio-Moreira2015compete,Walther-Franks2015robots,Williford2019framework,Wuertz2019healthy}

\nocite{adinolf2020little,aldemir2018qualitative,alexandrovsky2021serious,alexandrovsky2020playful,allart2017difficulty,allen2018satisfaction,altmeyer2020hexarcade,attig2019track,azadvar2018upeq,back2017designing}

\nocite{baek2017exploring,barata2017studying,blansonhenkemans2017design,bonus2015influence,born2021motivating,bowman2015when,burgers2015feedback,camingue2020visual,canossa2015impact,chen2016influence}

\nocite{chen2016scaffolding,cruz2017need,daneels2018enjoyment,degrove2015young,dechering2018moral,denden2021effects,denisova2020measuring,deterding2015joys,dindar2014motivational,ding2019applying}

\nocite{donnermann2021social,dunbar2014implicit,dunbar2018reliable,durga2015investigating,el-nasr2015unpacking,evin2020enabling,fairclough2020computer,fatehi2019gamifying,featherstone2019unicraft,feng2018gamification}

\nocite{frommel2021struggle,goh2017perceptions,gopinathbharathi2016knowledge,gray2019brainquest,gray2021space,groening2019achievement,gundry2018intrinsic,hamari2017why,hamlen2011childrens,hammad2021homecoming}

\nocite{hawlitschek2017increasing,herodotou2014dispelling,hietajarvi2019beyond,ho2017escaping,hosein2019girls,huang2011evaluating,hudson2016effects,jahn2021gamification,jang2016application,johnson2016motivations}

\nocite{johnson2018validation,johnson2021videogame,johnson2021need,kakoschke2021brain,kao2021effects,kao2021evaluating,kayali2018design,kim2015sense,kim2017mobile,king2018motivational}

\nocite{kleinman2021gang,koulouris2020effects,kumari2018why,landers2019greatest,li2015modeling,li2016player,liao2020impacts,lopez2019effects,lu2015kind,luu2017games}

\nocite{martinovic2016computer,mcgloin2016social,mckernan2015badges,mekler2017towards,merikivi2017what,miller2019expertise,mills2020self,morschheuser2017how,morschheuser2019cooperation,nagle2021pathfinder}

\nocite{nebel2016higher,neys2014exploring,nielsen2021teaching,oshea2019game,orji2017improving,palomaki2021link,partala2011psychological,passmore2020cheating,pe-than2014making,pedro2015badge}

\nocite{poeller2021prepare,poeller2021seek,poppelaars2018impact,rapp2017designing,ratan2015leveling,reer2018psychological,rieger2014winner,robinson2020designing,rodrigues2016ease,rodrigues2016playing}

\nocite{rodrigues2021personalisation,roest2015engaging,rogers2017motivational,roohi2020predicting,roy2016competitively,sailer2017how,saksono2020designing,salovaara2021programmable,sarkar2018meet,sarkar2018comparing}

\nocite{schimmenti2017schizotypal,schubhan2020investigating,shoshani2021fortnite,snodgrass2019towards,song2013effects,spann2019productive,steinberger2017road,sweetser2020game,tamplinwilson2019video,tinati2017investigation}

\nocite{tondello2019empirical,trang2021perils,tsai2021running,turkay2021virtual,tyack2021offpeak,tyack2021small,tyack2020restorative,vahlo2020challenge,vanhoudt2020disambiguating,vanroy2019unravelling}

\nocite{vanroy2019collecting,vandenabeele2020development,wang2011understanding,webster1993dimensionality,weech2020narrative,wiebe2014measuring,wiley2020points,yang2017examining,yucel2019battling}

\nocite{burnell2012breaking,cox2018good,griesemer2010design,griesemer2011design,griesemer2012design,hodent2014developing,hodent2015gamer,hodent2016gamer,hodent2017gamer3,hodent2019developing,hudson2011player,isaksen2012pursuit,lewis-evans2017throwing,rigby2008sustaining,shirinian2014video,yoder2018holy}

\bibliographystyle{ACM-Reference-Format}
\bibliography{sdt_tochi-acmsmall}

* Reference included in original corpus \citep[see also][]{tyack2020self}

** Reference included in expanded corpus 

\# Reference included in GDC corpus

\end{document}